\documentclass[%
reprint,
 amsmath,amssymb,
 prl,
floatfix,superscriptaddress,
]{revtex4-2}

\usepackage{multirow}
\usepackage{mathtools}
\usepackage{algorithmic}
\usepackage{algorithm}
\usepackage{float}
\usepackage{braket}
\usepackage{graphicx}
\usepackage{dcolumn}
\usepackage{bm}
\usepackage{inputenc}
\usepackage{blindtext}
\usepackage{color}
\usepackage{booktabs,siunitx}
\usepackage{hyperref}
\hypersetup{colorlinks,citecolor=blue}

\graphicspath{{Figure/}}
\newcolumntype{~}{>{\global\let\currentrowstyle\relax}}
\newcolumntype{^}{>{\currentrowstyle}}

\begin{document}

\newcommand{\dint}{\ensuremath{\mathrm{d}}}
\newcommand{\rvec}{\ensuremath{\mathbf{r}}}
\newcommand{\Rvec}{\ensuremath{\mathbf{R}}}
\newcommand{\epsilonvec}{\ensuremath{\pmb\epsilon}}
\newcommand{\etavec}{\ensuremath{\pmb\eta}}
\newcommand{\sigmavec}{\ensuremath{\pmb\sigma}}
\newcommand{\uvec}{\ensuremath{\mathbf{u}}}
\newcommand{\Pvec}{\ensuremath{\mathbf{P}}}
\newcommand{\Fvec}{\ensuremath{\mathbf{F}}}
\newcommand{\Wvec}{\ensuremath{\mathbf{W}}}
\newcommand{\Xvec}{\ensuremath{\mathbf{X}}}
\newcommand{\Yvec}{\ensuremath{\mathbf{Y}}}
\newcommand{\Xd}{\ensuremath{\widetilde{D}}}
\newcommand{\Xmlip}{\ensuremath{\widetilde{\pmb{D}}}}
\newcommand{\gammavec}{\ensuremath{\pmb{\gamma}}}
\newcommand{\Vmlip}{\ensuremath{\widetilde{V}_{\gammavec}}}
\newcommand{\qmlip}{\ensuremath{\widetilde{q}_{\gammavec}}}
\newcommand{\Nat}{\ensuremath{N_{\text{at}}}}
\newcommand{\Thetavec}{\ensuremath{\mathbf{\Theta}}}
\newcommand{\Zpart}{\ensuremath{\mathcal{Z}}}
\newcommand{\Fe}{\ensuremath{\mathcal{F}}}
\newcommand{\FeGB}{\ensuremath{\widetilde{\Fe}_{\gammavec}}}
\newcommand{\FeTDEP}{\ensuremath{\widetilde{\Fe}_{TDEP}}}
\newcommand{\Ra}{\ensuremath{\mathbfcal{R}}}
\newcommand{\He}{\ensuremath{\widetilde{\mathrm{H}}}}
\newcommand{\Ha}{\ensuremath{\mathrm{H}}}
\newcommand{\kBT}{\ensuremath{\mathrm{k}_\mathrm{B}\mathrm{T}}}
\newcommand{\epsi}{\ensuremath{\pmb{\varepsilon}}}
\newcommand{\DF}{\ensuremath{\mathcal{D}_{\mathrm{F}}}}
\newcommand{\DKL}{\ensuremath{\mathcal{D}_{\mathrm{KL}}}}

\preprint{APS/123-QED}

\title{{\it Ab initio} Canonical Sampling based on Variational Inference}


\author{Alo\"is Castellano}
\affiliation{CEA, DAM, DIF, F-91297 Arpajon, France, and Universit\'e Paris-Saclay, CEA, Laboratoires des Mat\'eriaux en Conditions Extr\^emes, 91680 Bruy\`eres-le-Ch\^atel, France.}
\author{Fran\c{c}ois Bottin}
\affiliation{CEA, DAM, DIF, F-91297 Arpajon, France, and Universit\'e Paris-Saclay, CEA, Laboratoires des Mat\'eriaux en Conditions Extr\^emes, 91680 Bruy\`eres-le-Ch\^atel, France.}
\author{Johann Bouchet}
\affiliation{CEA, DES, IRESNE, DEC, Cadarache, F-13018 St Paul Les Durance, France.}
\author{Antoine Levitt}
\affiliation{CERMICS, Ecole des Ponts,
Marne-la-Vallée, France \\ MATHERIALS team-project, Inria Paris, France}
\author{Gabriel Stoltz}
\affiliation{CERMICS, Ecole des Ponts,
Marne-la-Vallée, France \\ MATHERIALS team-project, Inria Paris, France}


\begin{abstract}
Finite temperature calculations, based on \textit{ab initio} molecular dynamics (AIMD) simulations, are a powerful tool able to predict material properties that cannot be deduced from ground state calculations.
However, the high computational cost of AIMD limits its applicability for large or complex systems.
To circumvent this limitation we introduce a new method named Machine Learning Assisted Canonical Sampling (MLACS), which accelerates the sampling of the Born--Oppenheimer potential surface in the canonical ensemble.
Based on a self-consistent variational procedure, the method iteratively trains a Machine Learning Interatomic Potential to generate configurations that approximate the canonical distribution of positions associated with the {\it ab initio} potential energy.
By proving the reliability of the method on anharmonic systems, we show that the method is able to reproduce the results of AIMD with an \textit{ab initio} accuracy at a fraction of its computational cost.

\end{abstract}

\maketitle






Molecular dynamics (MD) simulations, beside Monte-Carlo calculations~\cite{Metropolis_JCP21_1953}, are nowadays a popular way to obtain finite temperature properties and explore the phase diagram of materials. 
If the seminal works of Alder~\cite{Alder_JCP27_1957,Alder_JCP31_1959,Alder_JCP33_1960}, Rahman~\cite{Rahman_PR136_1964,Rahman_JCP55_1971} and coworkers only treated classical potentials, all the powerful tools introduced in these papers were reused in \textit{ab initio} molecular dynamics (AIMD) codes thereafter. 
In particular, by using the density functional theory (DFT)~\cite{Hohenberg_PR136_1964,Kohn_PR140_1965} and the Born--Oppenheimer (BO)~\cite{Born_Annalen20_1927} approximation, but also by assuming that the ground state electronic density is obtained at each MD time step~\cite{Kresse_PRB47_1993,Kresse_PRB48_1993}, all the ideas proposed previously apply. However, the high computational cost, due to both the evaluation of the Hellmann--Feynman forces using DFT and the high number of MD time steps required to sample the BO surface potential in the canonical ensemble, limits the range of systems that can be studied. To accelerate the computation of finite temperature properties, two main strategies have been proposed. In the first one, AIMD simulations are replaced by MD with \textit{ab initio}-based numerical potentials. In the second one, the sampling of the canonical distribution is performed through a direct generation of atomic configurations, bypassing AIMD simulations.

Recent progress in the field of Machine Learning Interatomic Potentials (MLIP)~\cite{Zuo2020, Behler2016, Bartk2010, Thompson2015, Novikov2021} promises an acceleration of finite-temperature studies with a near-DFT accuracy.
This high accuracy is ensured by the flexibility of MLIPs, which enables to reproduce a large variety of BO surfaces.
However, the construction of a MLIP is a tedious task as MLIP show poor extrapolative capabilities. 
Consequently, the set of configurations used for the MLIP requires a careful construction, which led to the development of learn-on-the-fly molecular dynamics~\cite{Podryabinkin2017,Li2015,Liu_PRM_2021} and other active learning based dataset selection methods~\cite{Dragoni2018,Deringer2017,Zhang2019}.
Moreover, usual finite-temperature works using MLIP shift the studied system from the DFT description to the MLIP one, with atomic positions (and computed properties) distributed according to the Boltzmann weights associated with the MLIP potential.
Consequently, one may need to ensure that simulations using MLIP are not in the extrapolative regime.
This verification can be done by computing corrections from free energy perturbation methods~\cite{Kruglov2019}, which however requires additional DFT single-point calculations. 

Several groups have recently proposed methods to bypass AIMD and generate configurations by adapting the Self-Consistent Harmonic Approximation (SCHA)~\cite{Gillis1968,Werthamer1970,Tadano2018,Esfarjani2020} to modern \textit{ab initio} methods, using a varational inference strategy.
Among those, we can cite the stochastic Temperature Dependent Effective Potential~\cite{Shulumba_PRB95_2017}, the Stochastic SCHA~\cite{Bianco2017,Monacelli2018,Monacelli2021} and the Quantum Self-Consistent Ab Initio Lattice Dynamics~\cite{vanRoekeghem2021}.
Within those methods (named EHCS in the following, for Effective Harmonic Canonical Sampling), configurations are generated with displacements around equilibrium positions according to a distribution corresponding to an effective harmonic Hamiltonian.
The self-consistent (SC) construction of this Hamiltonian, based on a variational procedure, allows to include explicit temperature effects.
However, the harmonic form of the effective potential means that the atomic displacements follow a Gaussian distribution.
This can entice differences for actual distributions of displacements in highly anharmonic solids or close to the melting temperature, and makes this approach completely inapplicable on liquids.

In this Letter, we propose a new method named Machine Learning Assisted Canonical Sampling (MLACS), which can be thought of as a generalization of the variational inference strategy used in the EHCS methods to linear MLIP.
MLACS consists in a SC variational procedure to generate configurations in order to best approximate the DFT canonical distribution and obtain a near-DFT accuracy in the properties computed at a fraction of the cost of AIMD.
In the present work, the MLIP is built to produce an effective canonical distribution which reproduces the DFT one at a single thermodynamical point, rather than sketching an effective BO potential surface for all thermodynamic conditions.
Therefore, MLACS is a sampling method which accelerates the computation of finite-temperature properties compared to AIMD simulations.


Let us consider an arbitrary system of $\Nat$ atoms (a crystal or a liquid, elemental or alloyed) at a temperature $T$, described by the coordinates $\Rvec \in \mathbb{R}^{3 \Nat}$. The potential energy $V(\Rvec)$ induces the canonical equilibrium distribution $p(\Rvec) = \mathrm{e}^{-\beta V(\Rvec)}/\Zpart$, with $\beta = 1/\mathrm{k_B T}$ and $\Zpart=\int \dint\Rvec \mathrm{e}^{-\beta V(\Rvec)}$ the partition function. For this system, the average of an observable $O(\Rvec)$, depending only on positions, is given by 
\begin{equation}
\label{eq:average}
    \braket{O} = \int \dint\Rvec O(\Rvec) p(\Rvec) 
               \approx \sum_n O(\Rvec_n) w_n \mbox{\; ,}
\end{equation}

\noindent with $w_n$ the weight of a configuration $n$, which follows the normalization $\sum_n w_n = 1$. Rather than performing AIMD simulations to generate a set of $\Rvec_n$ and $w_n$, the objective of MLACS is to generate a reduced set of configurations and weights using a MLIP and to compute them using DFT.

Let us define to this purpose a MLIP potential $\Vmlip(\Rvec) = \sum_{k=1}^K \Xd_k(\Rvec) \gamma_k$ with partition function $\widetilde{\Zpart}_{\gammavec}=\int \dint\Rvec \mathrm{e}^{-\beta \Vmlip(\Rvec)}$ and canonical equilibrium distribution $\qmlip(\Rvec)= \mathrm{e}^{-\beta \Vmlip(\Rvec)} / \widetilde{\Zpart}_{\gammavec}$.
The column vector $\gammavec \in \mathbb{R}^K$ contains adjustable parameters and the functions included in the row vector $\Xmlip : \mathbb{R}^{3\Nat}\rightarrow \mathbb{R}^K$ are named descriptors.
While a linear dependence is assumed between the potential energy and the descriptor space, no assumption is made on the form of $\Xmlip(\Rvec)$, enabling the use of a wide variety of descriptors, from a harmonic form to more sophisticated atom centered descriptors such as Atom Centered Symmetry Functions~\cite{Behler2011} or the Smooth Overlap of Atomic Positions~\cite{De2016,Bartk2010}. For the sake of simplicity and robustness, we use the Spectral Neighbor Analysis Potential~\cite{Thompson2015,Wood2018,Cusentino2020} in this work.

A key idea in MLACS is to use the distribution $\qmlip(\Rvec)$ instead of $p(\Rvec)$ in Eq.~\eqref{eq:average}.
With this approach, the acceleration is provided by the reduced computational cost to generate configurations using $\Vmlip(\Rvec)$ instead of $V(\Rvec)$.
To ensure accurate results, one needs to adjust the parameters $\gammavec$ of the effective potential so that $\qmlip(\Rvec)$ is the best approximation to the true distribution $p(\Rvec)$.
As a measure of the similarity between the two distributions $p$ and $\qmlip$, we use the Kullback--Leibler divergence (KLD)

\begin{equation}
\label{eq:KL}
    \DKL(\qmlip\Vert p) = \int \dint\Rvec \qmlip(\Rvec) \ln \left(\frac{\qmlip(\Rvec)}{p(\Rvec)}\right) \geq 0 \mbox{\; .}
\end{equation}

The KLD is a non-negative and asymmetric measure of the discrepancy between two distributions.
The smaller the KLD, the closer the distribution $p$ and $\qmlip$ are, with $\DKL(\qmlip\Vert p)=0$ meaning that the two distributions are identical.
So, by minimizing this quantity with respect to the parameters $\gammavec$, we obtain the distribution $\qmlip(\Rvec)$ that best reproduces $p(\Rvec)$.
A more tractable formulation can be obtain by transforming Eq.~\eqref{eq:KL} to an equivalent free energy minimisation (see SM I):

\begin{equation}
\label{eq:GB_fe}
    \Fe           \leq \min_{\gammavec} \; \FeGB \mbox{\; with \;}
    \FeGB = \FeGB^0 + \braket{V(\Rvec) - \Vmlip(\Rvec)}_{\Vmlip} \mbox{,}
\end{equation}

\noindent where $\braket{O}_{\Vmlip} = \int \dint \Rvec O(\Rvec) \qmlip(\Rvec)$ is the canonical average for the potential $\Vmlip(\Rvec)$, $\Fe = -\mathrm{k_B T} \ln(\Zpart)$ the free energy associated to $V(\Rvec)$, and $\FeGB^0 = -\mathrm{k_B T} \ln(\widetilde{\Zpart}_{\gammavec})$ the free energy associated to $\Vmlip(\Rvec)$. The Gibbs--Bogoliubov (GB) free energy $\FeGB$ written in Eq.~\eqref{eq:GB_fe} is the starting point for various variational procedures, in particular the SCHA~\cite{Gillis1968,Tadano2018,Monacelli2018}. By minimizing Eq.~\eqref{eq:GB_fe} with respect to $\gammavec$, we obtain (see SM II)

\begin{equation}
\label{eq:LS}
        \gammavec = \Braket{ \Xmlip(\Rvec)^\mathrm{T} \Xmlip(\Rvec)}_{\Vmlip}^{-1} \Braket{\Xmlip(\Rvec)^\mathrm{T} V(\Rvec)}_{\Vmlip} \mbox{\; ,}
\end{equation}
a non-trivial least-squares solution showing a circular dependency over $\gammavec$, solved using a SC variational procedure described below. Thus, the initial problem turns into an optimisation problem, solved with a flexible and simple MLIP potential, using variational inference~\cite{Blei_2017,Yang_ArXiv_2019}. So far, only the DFT data $V(\Rvec)$ are used, and not their derivatives. In order to lower the number of DFT calculations needed we also use the Fisher divergence~\cite{Yang_ArXiv_2019,Lyu_Proc_2009} $\DF(\qmlip \Vert p) = \int \dint\Rvec \qmlip(\Rvec) \vert \nabla_{\Rvec} \ln [ {\qmlip(\Rvec)}/{p(\Rvec)} ] \vert^2 \geq 0$, which measures the difference between two distributions using information on the gradients. The Fisher divergence can be rewritten (see SM III) as $\DF(\qmlip \Vert p) = \beta^2\braket{\vert \Fvec - \widetilde{\Fvec}_{\pmb{\gamma}} \vert^2}_{\Vmlip} \geq 0$, with $\Fvec=-\nabla_{\Rvec}V(\Rvec)$ and $\widetilde{\Fvec}_{\pmb{\gamma}}=-\nabla_{\Rvec}\Vmlip(\Rvec)$, the DFT and MLIP forces, respectively. Finally, the optimal parameters of the MLIP are obtained by minimizing the cost function $\Delta=\alpha_{\rm KL}\DKL + \alpha_{\rm F}\DF$ with respect to $\gammavec$, with $\alpha_{\rm KL}$ and $\alpha_{\rm F}$ the weights of each contribution. 

\begin{figure}
\includegraphics[scale = 0.12]{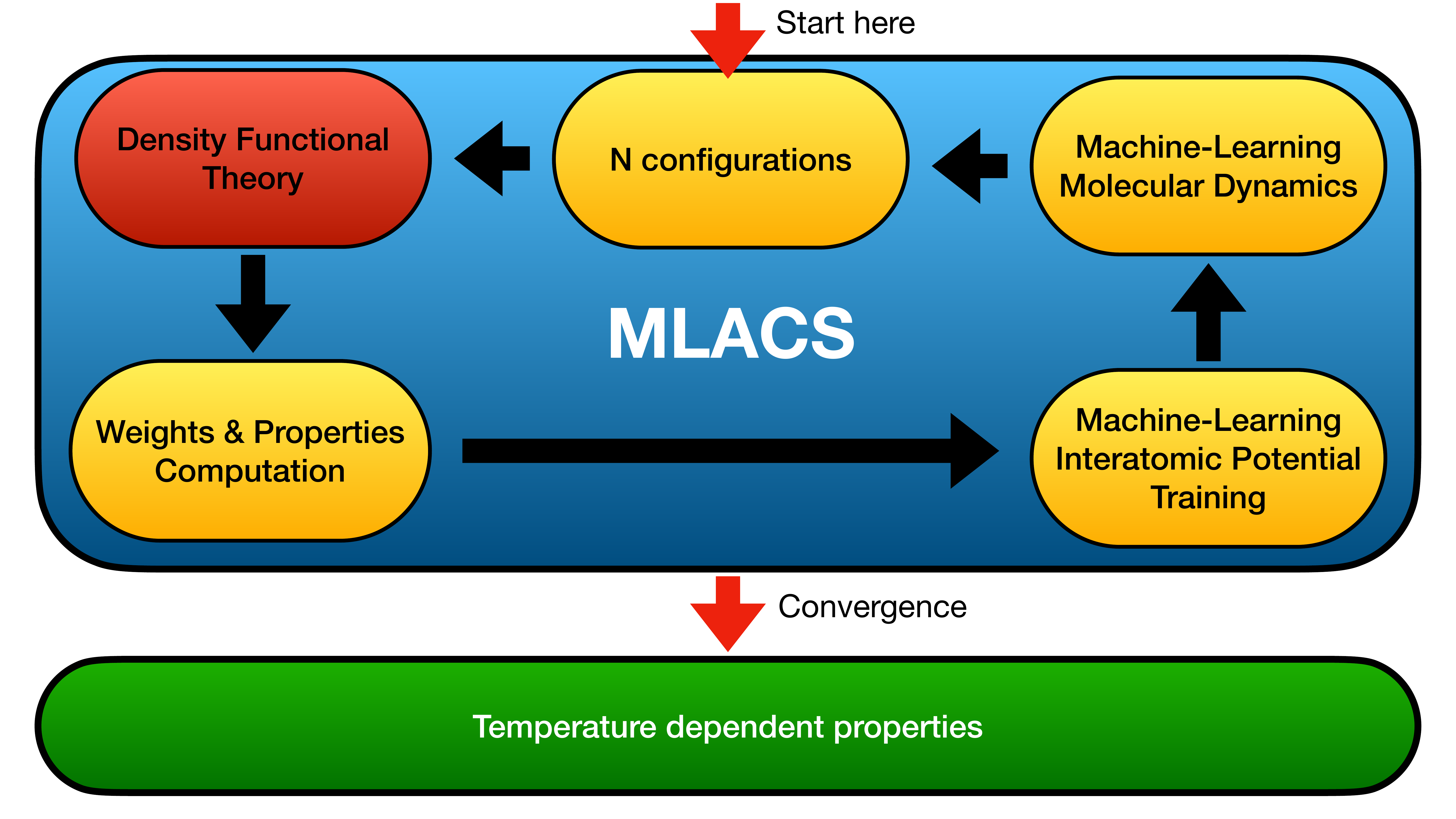}
\caption{\label{fig:MLACS} Workflow of MLACS. All the computational details are given in SM IV.}
\end{figure}

In order to perform this minimization, we adopt the following SC procedure (see Fig.~\ref{fig:MLACS}): start with $N$ atomic configurations, compute their energies, forces and stresses at the DFT level, check the convergence of specific observables (phonon spectrum, pair distribution function...), compute $\gammavec$, build a MLIP potential $\Vmlip$, perform a MD run using the MLIP (i.e. sample using $\Vmlip$) and extract new $N$ atomic configurations from the trajectory. This SC procedure is repeated until convergence.

Formally, the least-squares fitting has to be made using data distributed according to the last $\Vmlip$. To still reduce the number of DFT calculations, all the configurations from previous iterations are reused and reweighted (see SM IV) using the Multistate Bennett Acceptance Ratio (MBAR)~\cite{Shirts2008,shirts2017reweighting}. The weights $w_n$ obtained by this procedure are used to fit the MLIP potential and to evaluate physical properties (see Eq.~\eqref{eq:average}). Therefore, all observables computed using DFT (forces, energies and stresses, but also electronic properties) can be averaged at no extra cost, in contrast with usual MLIPs. Moreover, the free energy of the DFT system can be computed using both thermodynamic perturbation~\cite{Freitas_2016,Leite_2019} and cumulant expansions~\cite{Stoltz_2010}, without extra calculations (see SM V).

Concerning the computational cost, the acceleration enabled by MLACS with respect to AIMD calculations is twofold. First, MLACS strongly reduces the time required to generate independent configurations (around a hundred DFT calculations are needed and no longer thousands as in AIMD). Secondly, the calculation of the $N$ configurations can be parallelized at each iteration. Thus, an acceleration factor of 50 in computation time and 1000 in human time can be reached with $N=20$. Moreover, an additional acceleration can be obtained by using, as a starting point of MLACS, a MLIP potential coming from a previous calculation.


To demonstrate the accuracy and versatility of the method, we compare AIMD, MLACS and EHCS results on eight different systems (crystals, liquid and alloy) by performing three classical simulations (see SM VII) and five DFT calculations (see SM VIII). The {\it ab initio} calculations are performed using the ABINIT code~\cite{Gonze2020} over thousands of processors~\cite{Bottin_CMS_2007}. As a stringent test, we focus on the phonon spectrum, because it is particularly sensitive to the quality of the sampling so that its convergence guarantees the convergence of thermodynamical and elastic properties. These data are extracted using the Temperature Dependent Effective Potential (TDEP) method~\cite{Hellman_PRB84_2011} as implemented in ABINIT~\cite{Bottin_CPC_2020}. 

As a first DFT example, we consider the diamond phase of Si with a ($3\times3\times3$) supercell including 216 atoms at $\mathrm{T}=900~\mathrm{K}$. 
It has been shown using EHCS that anharmonicity is manifest at this temperature~\cite{Kim2018,Kim2020}, whereas Si is more harmonic at lower temperatures~\cite{Knoop2020}.
\begin{figure}
    \centering
    \includegraphics[width=\linewidth]{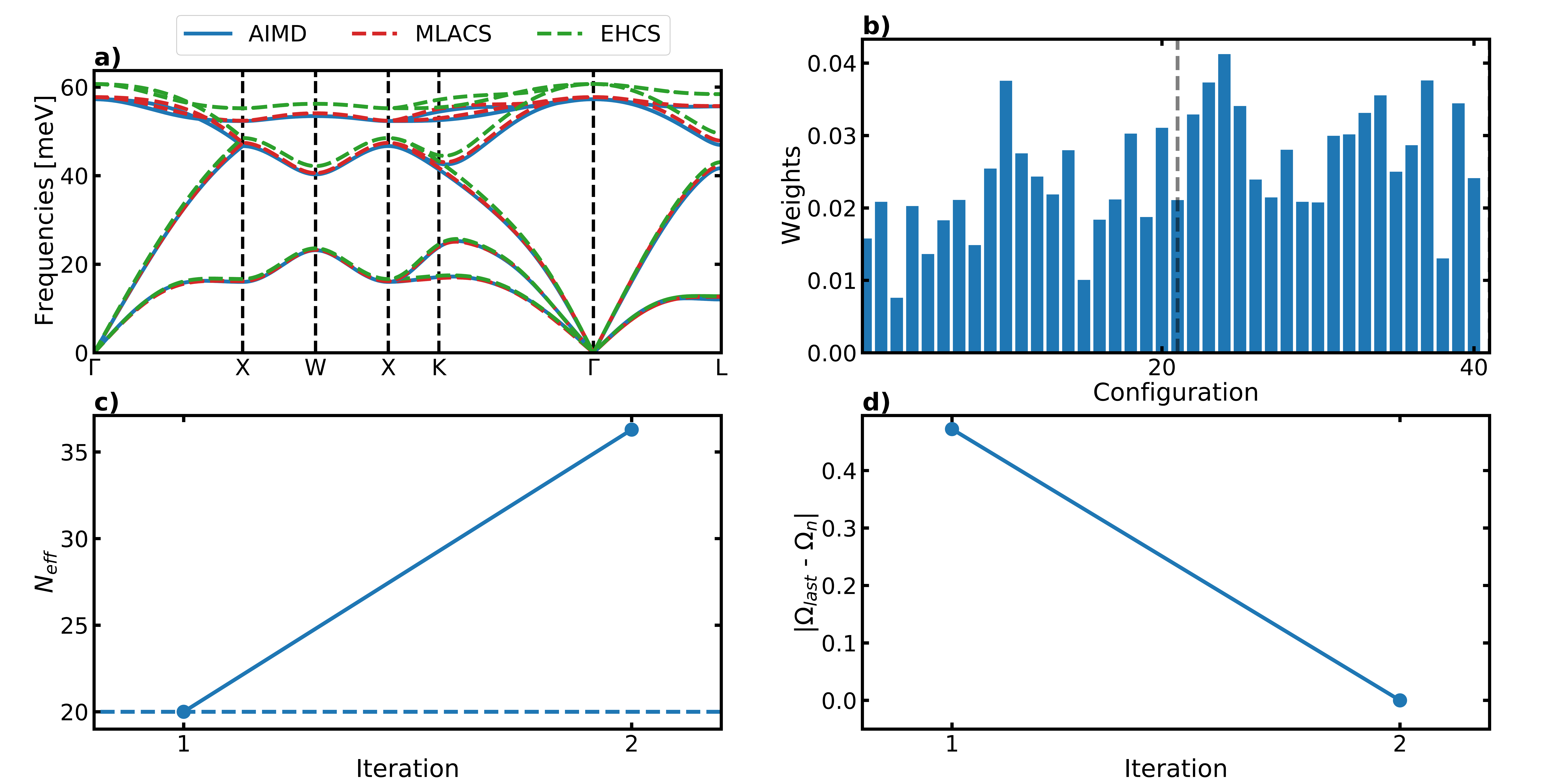}
    \caption{Results for Si at $\mathrm{T}=900~\mathrm{K}$. Main: TDEP phonon spectra for AIMD (blue), MLACS (red) and EHCS (green) simulations. Inset: weight of each configuration for the MLACS simulation.}
    \label{fig:Compare Si}
\end{figure}
Even though MLACS only requires $N_{\rm confs}= 60$ configurations (20 for the initialization and 40 for the production), the phonon spectrum of Si (see Fig.~\ref{fig:Compare Si}) is in excellent agreement with the AIMD one ($N_{\rm confs}= 3546$ during 9 ps). In contrast, the EHCS simulations ($N_{\rm confs}=$ 160) overestimate the optical branches by several meV. This fast and good convergence comes from both the ability of MLACS to generate uncorrelated configurations and the efficiency of the reweighting. Consequently, a near-DFT accuracy associated with a strong acceleration of the computational time, from one week with AIMD to 3 hours using MLACS, can be achieved.  

\begin{figure}[b]
    \centering
    \includegraphics[width=\linewidth]{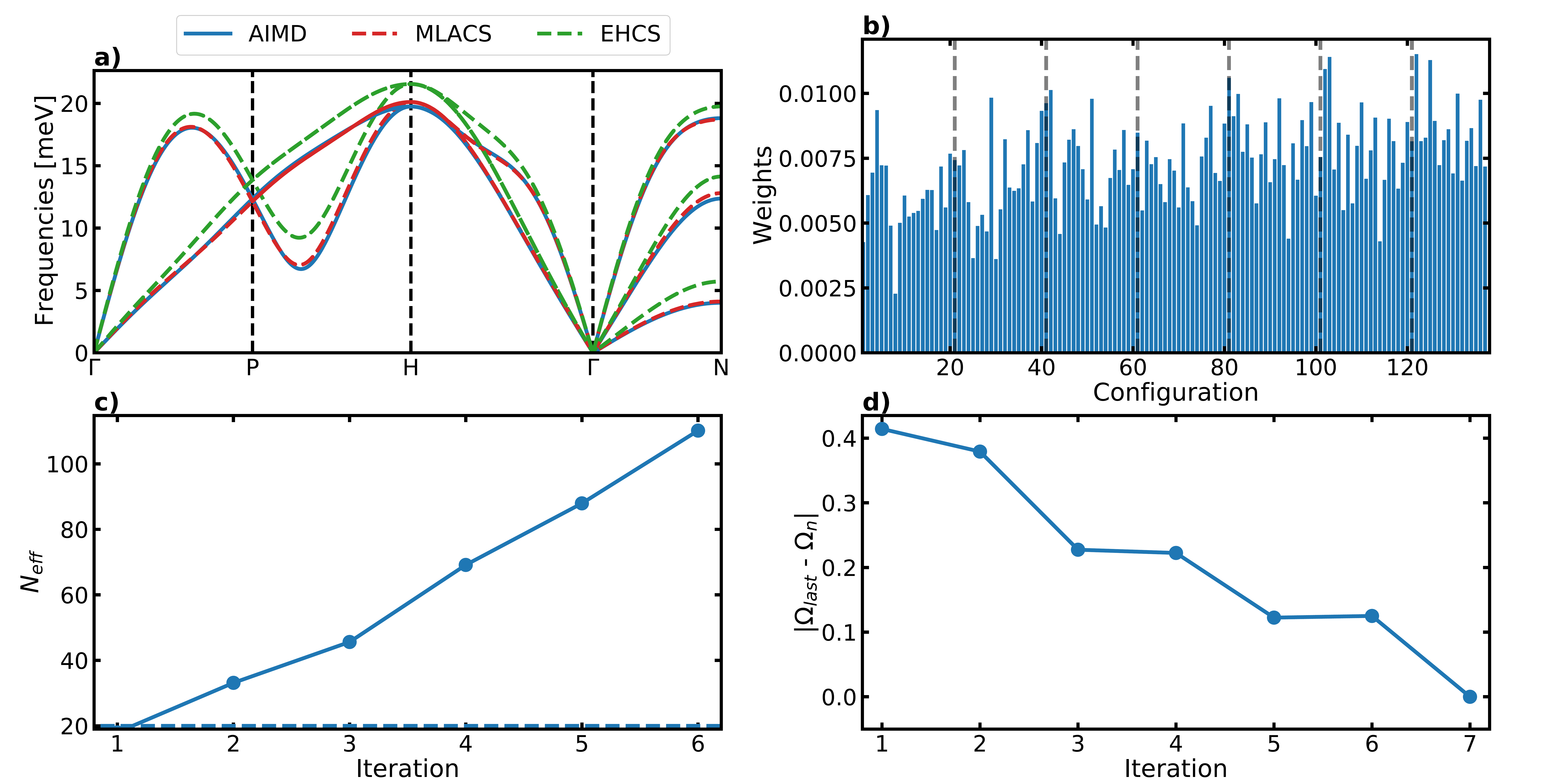}
    \caption{Same caption as Fig.~\ref{fig:Compare Si} for $\beta$-Zr at $\mathrm{T}=1000~\mathrm{K}$.}
    \label{fig:Compare Zr}
\end{figure}
\begin{figure*}[t]
    \centering
    \includegraphics[width=\linewidth]{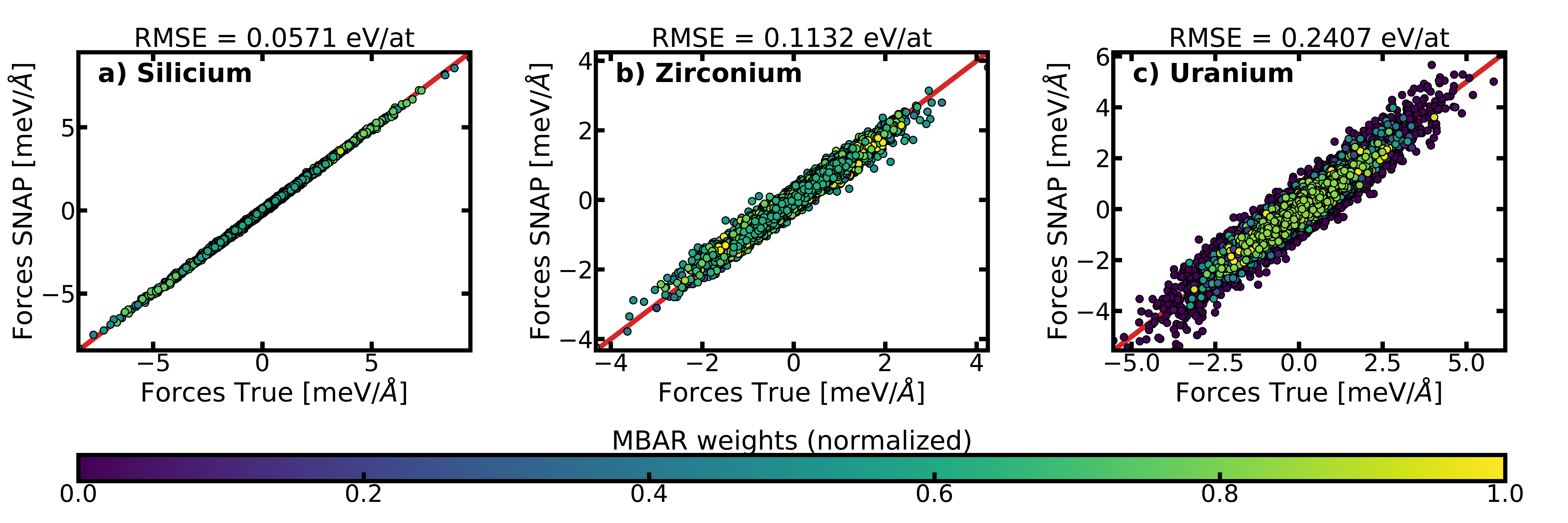}
    \caption{Correlations between AIMD and MLACS forces: a) Si at $\mathrm{T}=900~\mathrm{K}$, b) $\beta$-Zr at $\mathrm{T}=1000~\mathrm{K}$ and c) $\gamma$-U at $\mathrm{T}=1200~\mathrm{K}$.}
    \label{fig:Fit}
\end{figure*}
We next consider Zr in its bcc $\beta$ phase, with a $(4\times4\times4)$ supercell containing 128 atoms, at $\mathrm{T}=1000~\mathrm{K}$. 
The $\beta$-Zr phase being unstable at low temperatures, the stabilisation at higher temperatures comes from strong anharmonic effects~\cite{Hellman_PRB84_2011,Bottin_CPC_2020,Anzellini_2020}.
As previously, EHCS overestimates the whole phonon spectrum compared to AIMD (see Fig.~\ref{fig:Compare Zr}). Conversely, the phonon spectrum computed using MLACS is in excellent agreement with the AIMD one. Here, the acceleration in computational time is from two months with AIMD (with $N_{\rm confs}=$ 7758 during 19 ps) to two days using MLACS ($N_{\rm confs}=$ 160).

The third DFT example is the bcc $\gamma$ phase of Uranium with a $(4\times4\times4)$ supercell containing 128 atoms, at $\mathrm{T}=1200~\mathrm{K}$. 
As for $\beta$-Zr, $\gamma$-U is unstable at low temperature and is stabilized by anharmonic effects~\cite{Sderlind2012,Bouchet2017,Castellano2020}. The EHCS calculations never converged, despite several attempts and starting points (especially using the MLIP fitted using AIMD simulations).
\begin{figure}[b]
    \centering
    \includegraphics[width=\linewidth]{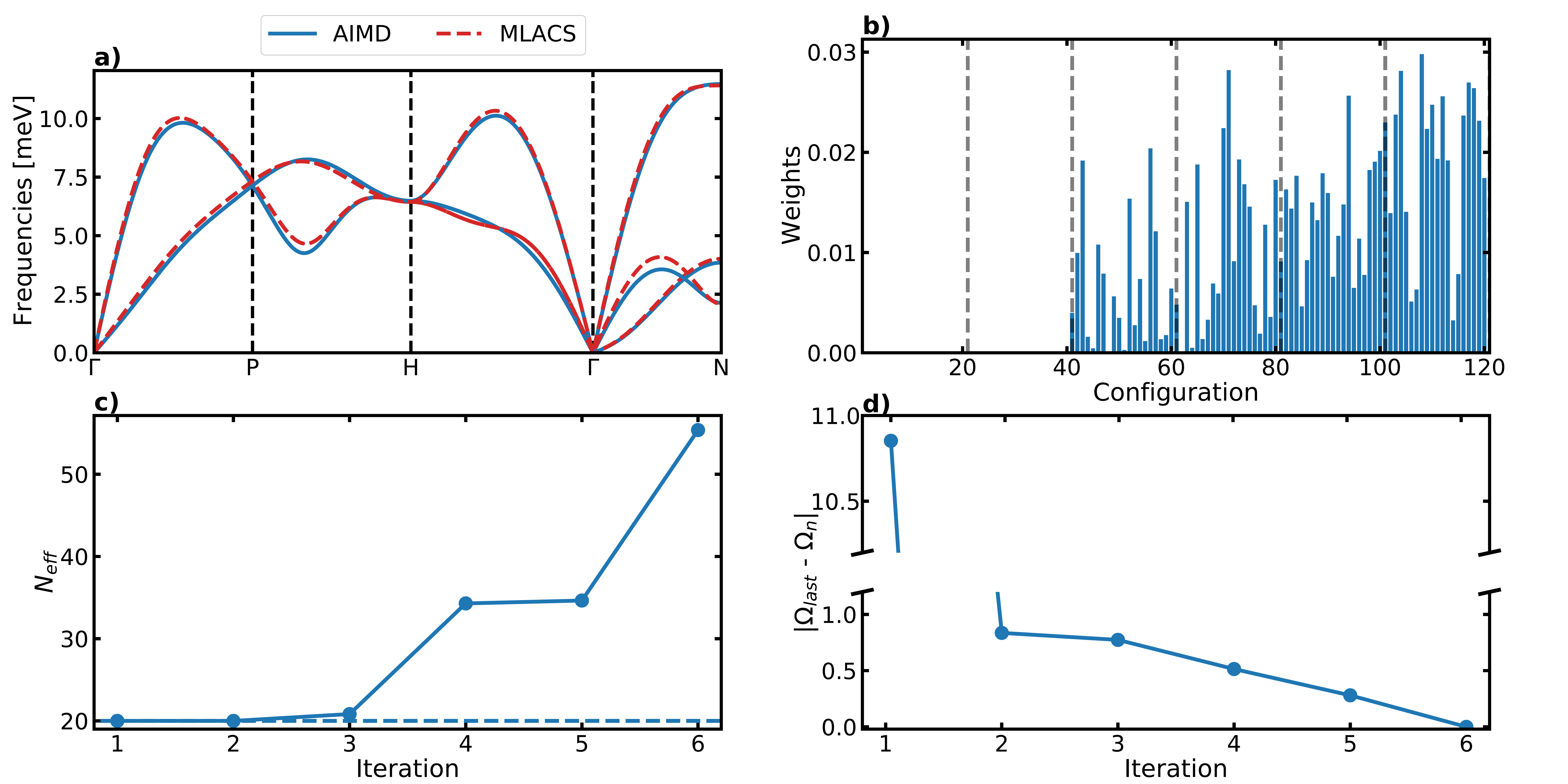}
    \caption{Same caption as Fig.~\ref{fig:Compare Si} for $\gamma$-U at $\mathrm{T}=1200~\mathrm{K}$.}
    \label{fig:Compare U}
\end{figure}
The phonon spectrum obtained using MLACS (see Fig.~\ref{fig:Compare U}) reproduces correctly the AIMD one, even if some small discrepancies remain (lower than 1 meV). 
Here, the reweighting is crucial, since the first 40 atomic configurations depart from the bcc phase (they look like a glass) and are discarded by MBAR from the statistical averages (see Fig.~\ref{fig:Fit}c and the inset in Fig.~\ref{fig:Compare U}). After them, the next 80 ones get closer to the final structure and have a non-zero weight in the equilibrium canonical distribution. This differs strongly from Si (see Fig.~\ref{fig:Fit}a and the inset in Fig.~\ref{fig:Compare Si}) and $\beta$-Zr (see Fig.~\ref{fig:Fit}b and the inset in Fig.~\ref{fig:Compare Zr}), for which all the configurations almost equally contribute. Despite this loss of data, MLACS still leads to a significant reduction in the computational cost from two months with AIMD ($N_{\rm confs}=$ 5981 during 22 ps) to two days using MLACS ($N_{\rm confs}=$ 140).

\begin{figure}
    \centering
    \includegraphics[width=\linewidth]{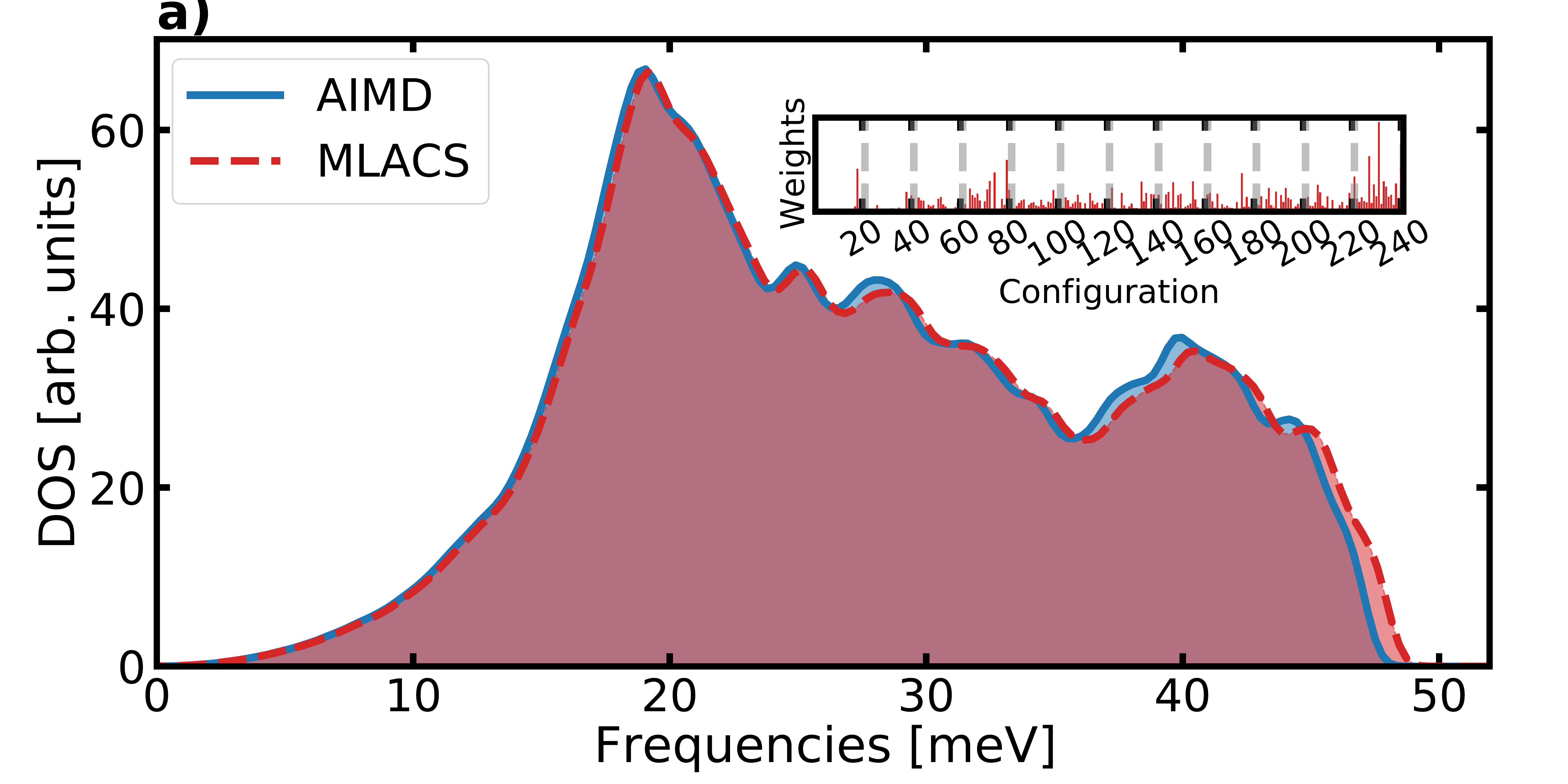}
    \caption{Phonon density of states of Al$_{0.5}$Cu$_{0.5}$ at $\mathrm{T}=600~\mathrm{K}$ obtained using MLACS and MD simulations.}
    \label{fig:AlCu}
\end{figure}
\begin{figure}
    \centering
    \includegraphics[width=\linewidth]{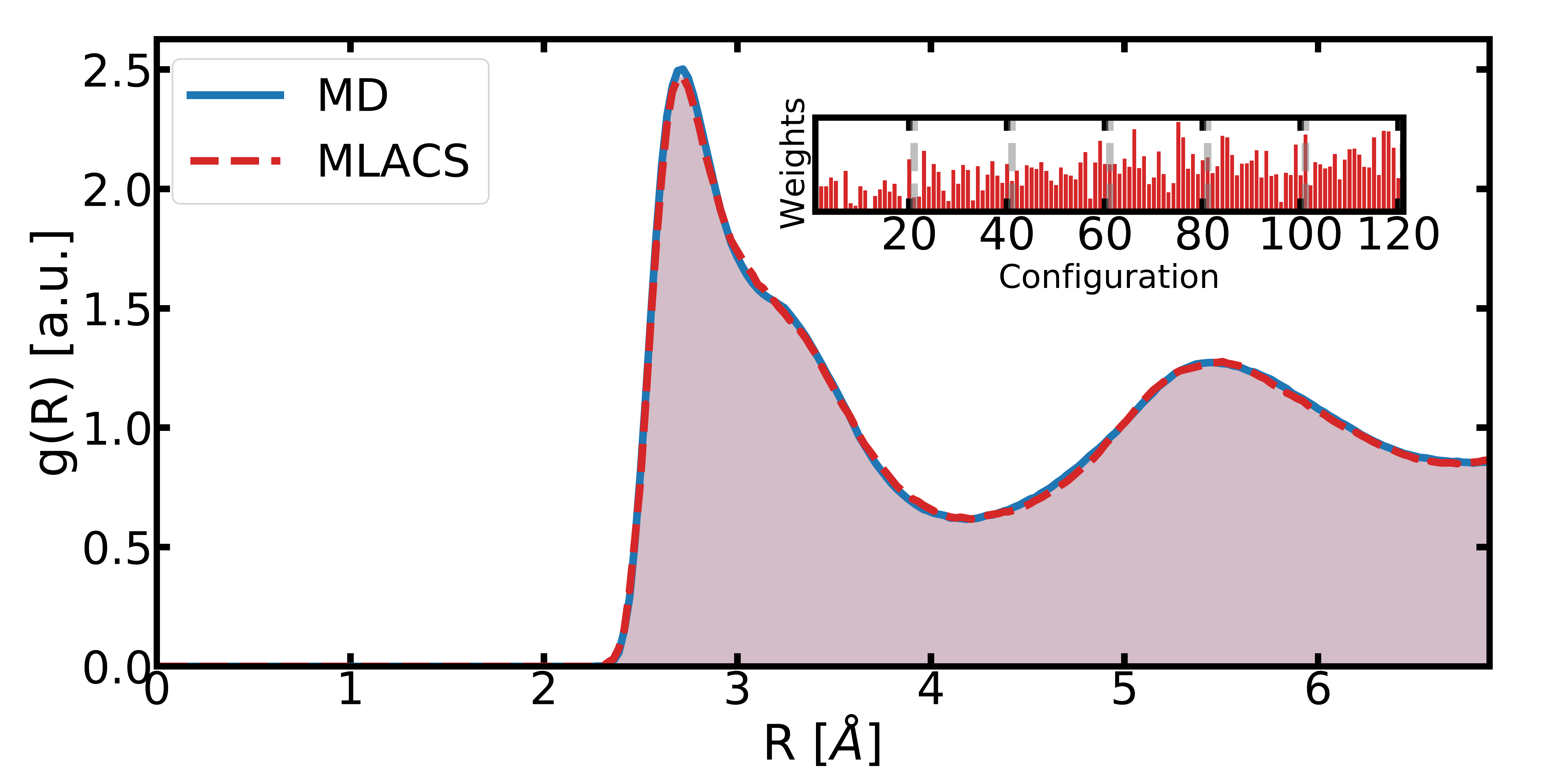}
    \caption{Pair distribution function of U liquid at $\mathrm{T}=2500~\mathrm{K}$ obtained using MLACS and MD simulations.}
    \label{fig:Uliquid}
\end{figure}
We also present three systems simulated with classical potentials and with sufficiently long MD trajectories (including tens of thousands time steps), which would be impossible to simulate using DFT. The first one is Silicon at $\mathrm{T}=1500~\mathrm{K}$ using a Tersoff potential, the second one is the Al$_{0.5}$Cu$_{0.5}$ alloy at $\mathrm{T}=600~\mathrm{K}$ using an Angular-Dependent Potential (ADP), and the third one is the liquid phase of Uranium at $\mathrm{T}=2500~\mathrm{K}$, using a Modified Embedded-Atom Method (MEAM) potential. In Fig.~\ref{fig:AlCu} and~\ref{fig:Uliquid}, we show that both the phonon density of states (DOS) of Al$_{0.5}$Cu$_{0.5}$ and the pair distribution function (PDF) of U liquid obtained using MLACS are in excellent agreement with the ones obtained using MD simulations, respectively. We stress that alloys and liquids cannot be computed using EHCS. 

In addition, we also compare the reference $\Fe$ and GB $\FeGB$ free energies for these three systems described by empirical potentials (see Tab.~\ref{tab:Compare SiZrU} and Eq.~\eqref{eq:GB_fe}). They agree up to an error lower than 1 meV, showing that the GB minimization has been effectively performed. 
Even if these results are obtained using classical simulations (thermodynamic integration requires hundreds of thousands MD time steps), they pave the way for performing free energy calculations with a near-DFT accuracy at a computational cost equivalent to one hundred single-point DFT calculations. Thus, a good approximation of the high temperature free energy is not only reachable for crystals dynamically stable at $\mathrm{T}=0~\mathrm{K}$~\cite{Glensk_PRL114_2015}, but also for strongly anharmonic systems such as $\beta$-Zr or $\gamma$-U and liquids.       

\begin{table}
\centering
    \caption{\label{tab:Compare SiZrU} Free energy $\Fe$ of the reference system compared to the GB one $\FeGB=\FeGB^0+\Delta\Fe$ for Si at $\mathrm{T}=2500~\mathrm{K}$, Al$_{0.5}$Cu$_{0.5}$ at $\mathrm{T}=600~\mathrm{K}$ and U liquid at $\mathrm{T}=2500~\mathrm{K}$. $\Fe$ is computed using a direct thermodynamic integration, whereas $\FeGB$ is evaluated using a thermodynamic integration giving $\FeGB^0$ and adding a correction $\Delta\Fe$ from a cumulant expansion (see SM V).}
\begin{tabular}{c c c c c }
\hline \hline \\ [-0.3cm]
System               & $\FeGB^0$ (eV/at) & $\FeGB$ (eV/at) & $\Fe$ (eV/at)\\
\hline
Silicon              & -5.0846           & -5.0845         & -5.0843  \\
Al$_{0.5}$Cu$_{0.5}$ & -3.7044           & -3.7076         & -3.7073  \\
Uranium liquid       & -7.4759           & -7.4786         & -7.4790 \\
\hline \hline
\end{tabular}
\end{table} 


In summary, we introduced MLACS, a method to sample the canonical ensemble and accelerate the {\it ab initio} computation of finite-temperature properties. By combining a self-consistent variational procedure with the highly flexible forms of MLIP, MLACS is able to include the effect of temperature with a near-DFT accuracy at a fraction of the cost of AIMD. We demonstrated the ability of the method to describe the thermodynamic and elastic properties of highly anharmonic crystals, liquids and alloys, in contrast to EHCS methods. We think that MLACS may help to more efficiently build the large databases needed to fit neural network models~\cite{Zhang_PRL120_2018}. Besides its reliability and the computational gain it provides, MLACS could be used to study systems whose complexity prevents the use of AIMD or surpasses the number of electrons which can be simulated using AIMD. In the future, we also believe that the MLACS strategy can be generalized to {\it ab initio} path-integral MD simulations in order to reduce the high computational cost of these calculations, while capturing the anharmonicy inherent to the quantum fluctuations of light elements.


\appendix


\bibliography{article}
\bibliographystyle{apsrev4-2}
\end{document}


\newcommand{\dint}{\ensuremath{\mathrm{d}}}
\newcommand{\rvec}{\ensuremath{\mathbf{r}}}
\newcommand{\Rvec}{\ensuremath{\mathbf{R}}}
\newcommand{\epsilonvec}{\ensuremath{\pmb\epsilon}}
\newcommand{\etavec}{\ensuremath{\pmb\eta}}
\newcommand{\sigmavec}{\ensuremath{\pmb\sigma}}
\newcommand{\alphavec}{\ensuremath{\pmb\alpha}}
\newcommand{\muvec}{\ensuremath{\pmb\mu}}
\newcommand{\uvec}{\ensuremath{\mathbf{u}}}
\newcommand{\Pvec}{\ensuremath{\mathbf{P}}}
\newcommand{\Fvec}{\ensuremath{\mathbf{F}}}
\newcommand{\Wvec}{\ensuremath{\mathbf{W}}}
\newcommand{\Xvec}{\ensuremath{\mathbf{X}}}
\newcommand{\Yvec}{\ensuremath{\mathbf{Y}}}
\newcommand{\Xd}{\ensuremath{\widetilde{D}}}
\newcommand{\Xmlip}{\ensuremath{\widetilde{\pmb{D}}}}
\newcommand{\gammavec}{\ensuremath{\pmb{\gamma}}}
\newcommand{\Vmlip}{\ensuremath{\widetilde{V}_{\gammavec}}}
\newcommand{\qmlip}{\ensuremath{\widetilde{q}_{\gammavec}}}
\newcommand{\Nat}{\ensuremath{N_{\text{at}}}}
\newcommand{\Thetavec}{\ensuremath{\mathbf{\Theta}}}
\newcommand{\Zpart}{\ensuremath{\mathcal{Z}}}
\newcommand{\Fe}{\ensuremath{\mathcal{F}}}
\newcommand{\FeGB}{\ensuremath{\widetilde{\Fe}_{\gammavec}}}
\newcommand{\FeTDEP}{\ensuremath{\widetilde{\Fe}_{\mathrm{TDEP}}}}
\newcommand{\Ra}{\ensuremath{\mathbfcal{R}}}
\newcommand{\He}{\ensuremath{\widetilde{\mathrm{H}}}}
\newcommand{\Ha}{\ensuremath{\mathrm{H}}}
\newcommand{\kBT}{\ensuremath{\mathrm{k}_\mathrm{B}\mathrm{T}}}
\newcommand{\epsi}{\ensuremath{\pmb{\varepsilon}}}
\newcommand{\DF}{\ensuremath{\mathcal{D}_{\mathrm{F}}}}
\newcommand{\DKL}{\ensuremath{\mathcal{D}_{\mathrm{KL}}}}

\preprint{APS/123-QED}

\title{{\it Ab initio} Canonical Sampling based on Variational Inference \\ (Supplementary Material)}

\author{Alo\"is Castellano}
\affiliation{CEA, DAM, DIF, F-91297 Arpajon, France, and Universit\'e Paris-Saclay, CEA, Laboratoires des Mat\'eriaux en Conditions Extr\^emes, 91680 Bruy\`eres-le-Ch\^atel, France.}
\author{Fran\c{c}ois Bottin}
\affiliation{CEA, DAM, DIF, F-91297 Arpajon, France, and Universit\'e Paris-Saclay, CEA, Laboratoires des Mat\'eriaux en Conditions Extr\^emes, 91680 Bruy\`eres-le-Ch\^atel, France.}
\author{Johann Bouchet}
\affiliation{CEA, DES, IRESNE, DEC, Cadarache, F-13018 St Paul Les Durance, France.}
\author{Antoine Levitt}
\affiliation{CERMICS, Ecole des Ponts,
Marne-la-Vallée, France \\ MATHERIALS team-project, Inria Paris, France}
\author{Gabriel Stoltz}
\affiliation{CERMICS, Ecole des Ponts,
Marne-la-Vallée, France \\ MATHERIALS team-project, Inria Paris, France}

\begin{abstract}
In this supplementary material, we give mathematical proofs, implementation details and additional results which are not given in the main paper.
\begin{enumerate}
    \item From the Kullback--Leibler divergence to the Gibbs--Bogoliubov free energy
    \item From the Gibbs--Bogoliubov free energy to a least-squares solution
    \item From the Fisher divergence to a least-squares solution
    \item The MLACS procedure
    \item Free energy calculations
    \item Implementation of MLACS
    \item Classical simulations with MLACS, MD and EHCS
    \item {\it Ab initio} simulations with MLACS, AIMD and EHCS
\end{enumerate}

\begin{center}
\begin{ruledtabular}
\begin{tabular}{lc}
 AIMD & {\it Ab initio} Molecular Dynamics \\
 DFT & Density Functional Theory \\
 EHCS & Effective Harmonic Canonical Sampling \\
 MD & Molecular Dynamics \\
 MLIP & Machine Learning Interatomic Potential \\
 MLMD & Machine Learning Molecular Dynamics \\
 MLACS & Machine Learning Assisted Canonical Sampling \\
 TDEP & Temperature Dependent Effective Potential \\
 \end{tabular}
\end{ruledtabular}
\end{center}

\end{abstract}

\maketitle


As in the main paper, let us consider a ``reference" (DFT) system with $\Nat$ atoms, at a temperature T. On the other hand, we define a ``surrogate" (MLIP) system, which becomes closer the first as the iterations of the algorithm proceed. The potential energy, canonical equilibrium distribution, partition function, free energy and average of the ``reference" and ``surrogate" systems are listed in Table~\ref{tab:def}.

\begin{table*}
\caption{\label{tab:def} Definition of various quantities of the reference and surrogate systems.}
\begin{tabular}{c@{\hspace{2cm}}rl@{\hspace{2cm}}rl}
\hline \hline
System & \multicolumn{2}{l}{Reference (DFT)} & \multicolumn{2}{l}{Surrogate (MLIP)} \\ 
\hline \\
 Potential energy & $V(\Rvec)$ & & $\Vmlip(\Rvec)$ & $= \Xmlip(\Rvec) \gammavec$ \\ 
 & & & & $= \sum_{k=1}^K \Xd_k(\Rvec) \gamma_k$ \\[0.2cm]
 Canonical equilibrium distribution & $p(\Rvec)$ & $= \frac{\mathrm{e}^{-\beta V(\Rvec)}}{\Zpart}$ & $\qmlip(\Rvec)$ & $= \frac{\mathrm{e}^{-\beta \Vmlip(\Rvec)}}{\widetilde{\Zpart}_{\gammavec}}$ \\  [0.2cm]
 Partition Function & $\Zpart$ & $=\Omega_0^{-1}\int \text{d}\Rvec \mathrm{e}^{-\beta V(\Rvec)}$ & $\widetilde{\Zpart}_{\gammavec}$ & $=\Omega_0^{-1}\int \text{d}\Rvec \mathrm{e}^{-\beta \Vmlip(\Rvec)}$ \\ [0.2cm]
 Free energy & $\Fe $ & $= -\kBT \ln(\Zpart)$ & $\FeGB^0 $ & $= -\kBT \ln(\widetilde{\Zpart}_{\gammavec})$ \\ [0.2cm]
 Average of an observable $O(\Rvec)$ & $\braket{O} $ & $= \int\dint\Rvec O(\Rvec) p(\Rvec)$ & $\braket{O}_{\Vmlip} $ & $= \int \dint\Rvec O(\Rvec) \qmlip(\Rvec)$ \\ [0.2cm] 
\hline \hline
\end{tabular}
\end{table*}


\section{From the Kullback--Leibler divergence to the Gibbs--Bogoliubov free energy \label{sec:KLD}}
The aim of this section is to derive Eq.~(3) in the main text. This bound is derived in the Self-Consistent Harmonic Approximation (SCHA) method~\cite{Gillis1968,Werthamer1970,Bianco2017} by a convexity inequality based on the free energy equality~\eqref{eq:Fequality} below. We present here an alternative derivation relying on the Kullback--Leibler (KL) entropy, as done in the statistics literature on variational inference (where the counterpart of our free energy upper bound is known as ELBO). Recall the KL divergence $\DKL$, based on the Shannon entropy, which writes:
\begin{equation}
\label{eq:DKL}
        \DKL(\qmlip \Vert p) = \int \dint \Rvec \qmlip(\Rvec) \ln \bigg[ \frac{\qmlip(\Rvec)}{p(\Rvec)} \bigg] \geq 0 \mbox{\; .}
\end{equation}
By straightforward computations, $\DKL(\qmlip \Vert p)$ can be rewritten in terms of free energies:
\begin{equation}
\begin{split}
        &\DKL(\qmlip \Vert p) = \int \dint \Rvec \qmlip(\Rvec) \ln \bigg[ \frac{\qmlip(\Rvec)}{p(\Rvec)} \bigg] \\
                             &= \int \dint \Rvec \qmlip(\Rvec) \bigg( \ln \bigg[\frac{\Zpart}{\widetilde{\Zpart}_{\gammavec}}\bigg] + \ln \bigg[ \mathrm{e}^{-\beta \big( \Vmlip(\Rvec) - V(\Rvec) \big)} \bigg] \bigg) \\
                             &= \ln (\Zpart) - \ln(\widetilde{\Zpart}_{\gammavec}) + \beta \braket{V(\Rvec) - \Vmlip(\Rvec)}_{\Vmlip} \\
                             &=-\beta (\Fe -\FeGB)
\label{eq:DKL_GB}
\end{split}
\end{equation}
with $\FeGB = \FeGB^0 + \braket{V(\Rvec) - \Vmlip(\Rvec)}_{\Vmlip}$. Thus, the minimization of the KL divergence (see Eq.~\eqref{eq:DKL}), is equivalent to the Gibbs--Bogoliubov inequality:
\begin{equation}
\label{eq:GB}
    \Fe           \leq \min_{\gammavec} \; \FeGB \mbox{\; .}
\end{equation}
Conversely, minimizing the Gibbs--Bogoliubov free energy $\FeGB$ provides the best approximation of both the free energy $\Fe$ and the distribution $p(\Rvec)$.

\section{From the Gibbs--Bogoliubov free energy to a least-squares solution \label{sec:GB-LS}}
In this section we show how to find the $\gammavec$ parameters. As discussed in Section~\ref{sec:KLD}, the minimization of the KL divergence written in Eq.~\eqref{eq:DKL_GB} with respect to $\gammavec$ is equivalent to the minimization of the Gibbs--Bogoliubov free energy. The gradient of the former quantity is:
\begin{equation}
\nabla_{\gammavec} \DKL = -\beta \nabla_{\gammavec} (\Fe - \FeGB) = \beta \nabla_{\gammavec} \FeGB  \mbox{\; .}
\end{equation}
Consequently, we focus on the minimization of the GB free energy in the following and find out when its minimum is reached ($\nabla_{\gammavec} \FeGB = 0$). The gradient of the Gibbs--Bogoliubov free energy is: 
\begin{equation}
\label{eq:grad FeGB}
    \nabla_{\gammavec} \FeGB = \nabla_{\gammavec} \FeGB^0 + \nabla_{\gammavec} \bigg[\braket{V(\Rvec) - \Vmlip(\Rvec)}_{\Vmlip}\bigg] \mbox{\; .}
\end{equation}
The gradient of the first term corresponds to the average of the descriptors. Indeed, using $\nabla_{\gammavec} \Vmlip(\Rvec) = \Xmlip(\Rvec)^\mathrm{T}$:
\begin{equation}
\label{eq:grad F0}
\begin{split}
        \nabla_{\gammavec} \FeGB^0 &= -\kBT \nabla_{\gammavec} \bigg(\ln \bigg[ \int \dint \Rvec \mathrm{e}^{-\beta \Vmlip(\Rvec)} \bigg]\bigg) \\
                                   &= \kBT \frac{\int \dint \Rvec \beta \Xmlip(\Rvec)^\mathrm{T} \mathrm{e}^{-\beta \Vmlip(\Rvec)}}{\int \dint \Rvec \mathrm{e}^{-\beta \Vmlip(\Rvec)}} \\
                                   &= \braket{\Xmlip(\Rvec)^\mathrm{T}}_{\Vmlip} \mbox{\; .}
\end{split}
\end{equation}
For the second term on the right-hand side of Eq.~\eqref{eq:grad FeGB}, we start by writing the gradient of the average reference potential $V(\Rvec)$ with respect to the $\gammavec$ MLIP parameters: 
\begin{equation}
\label{eq:grad V}
\begin{split}
        \nabla_{\gammavec}& \bigg[\braket{V(\Rvec)}_{\Vmlip}\bigg] = \nabla_{\gammavec} \bigg[\frac{1}{\widetilde{\Zpart}_{\gammavec}} \int \dint \Rvec V(\Rvec) \mathrm{e}^{-\beta \Vmlip(\Rvec)}\bigg] \\
        =&  \nabla_{\gammavec} \bigg( \frac{1}{\widetilde{\Zpart}_{\gammavec}} \bigg) \int \dint \Rvec V(\Rvec) \mathrm{e}^{-\beta \Vmlip(\Rvec)} \\ 
        &+ \frac{1}{\widetilde{\Zpart}_{\gammavec}} \int \dint \Rvec V(\Rvec) \nabla_{\gammavec} \bigg(\mathrm{e}^{-\beta \Vmlip(\Rvec)}\bigg) \\
        =& \beta \bigg( \braket{\Xmlip(\Rvec)^\mathrm{T}}_{\Vmlip} \braket{ V(\Rvec)}_{\Vmlip} - \braket{ V(\Rvec)\Xmlip(\Rvec)^\mathrm{T}}_{\Vmlip} \bigg) \mbox{\; .}
\end{split}
\end{equation}
We also write the gradient of the average MLIP potential $\Vmlip(\Rvec)$ with respect to $\gammavec$:
\begin{equation}
\label{eq:grad Vmlip}
\begin{split}
        \nabla_{\gammavec}& \bigg[\braket{\Vmlip(\Rvec)}_{\Vmlip}\bigg] = \nabla_{\gammavec} \bigg[\frac{1}{\widetilde{\Zpart}_{\gammavec}} \int \dint \Rvec \Vmlip(\Rvec) \mathrm{e}^{-\beta \Vmlip(\Rvec)}\bigg] \\
        =&  \nabla_{\gammavec} \bigg( \frac{1}{\widetilde{\Zpart}_{\gammavec}} \bigg) \int \dint \Rvec \Vmlip(\Rvec) \mathrm{e}^{-\beta \Vmlip(\Rvec)} \\ 
        &+ \frac{1}{\widetilde{\Zpart}_{\gammavec}} \int \dint \Rvec \nabla_{\gammavec} \bigg(\Vmlip(\Rvec)\bigg) \mathrm{e}^{-\beta \Vmlip(\Rvec)}\\
        &+ \frac{1}{\widetilde{\Zpart}_{\gammavec}} \int \dint \Rvec \Vmlip(\Rvec) \nabla_{\gammavec} \bigg(\mathrm{e}^{-\beta \Vmlip(\Rvec)}\bigg) \\
        =& \beta \bigg( \braket{\Xmlip(\Rvec)^\mathrm{T}}_{\Vmlip} \braket{ \Vmlip(\Rvec)}_{\Vmlip} - \braket{ \Vmlip(\Rvec)\Xmlip(\Rvec)^\mathrm{T}}_{\Vmlip} \bigg) \\ 
        &+ \braket{\Xmlip(\Rvec)^\mathrm{T}}_{\Vmlip} \mbox{\; .}
\end{split}
\end{equation}
If we assume that the average of the reference potential equals zero, the term $\braket{\Xmlip(\Rvec)^\mathrm{T}}_{\Vmlip} = 0$ will have no contribution. Then, by injecting Eqs.~\eqref{eq:grad F0},~\eqref{eq:grad V} and~\eqref{eq:grad Vmlip} into Eq.~\eqref{eq:grad FeGB}, one obtains:
\begin{equation}
        \nabla_{\gammavec} \FeGB = -\beta \bigg[ \Braket{ \Xmlip(\Rvec)^\mathrm{T}V(\Rvec)}_{\Vmlip}  - \Braket{\big(\Xmlip(\Rvec)^\mathrm{T}\Vmlip(\Rvec)}_{\Vmlip} \bigg] \mbox{\; .} \nonumber
\end{equation}
By using the expression of the surrogate potential $\Vmlip(\Rvec)=\Xmlip(\Rvec)\gammavec$, which corresponds to write the vector components of $\gammavec = \begin{bmatrix} \gamma_1 , \gamma_2 , \ldots , \gamma_{k} \end{bmatrix}^{\mathrm{T}}$ and $\Xmlip = \begin{bmatrix} \Xd_1 , \Xd_2 , \ldots , \Xd_{k} \end{bmatrix}$ in column and row respectively, we find the parameters $\gammavec$ at the minimum of the Gibbs--Bogoliubov free energy ($\nabla_{\gammavec} \FeGB = 0$):
\begin{equation}
\label{eq:WLS}
        \gammavec = \Braket{\Xmlip(\Rvec)^\mathrm{T} \Xmlip(\Rvec)}_{\Vmlip}^{-1}
                   \Braket{ \Xmlip(\Rvec)^\mathrm{T}  V(\Rvec)}_{\Vmlip}  \mbox{\; ,}
\end{equation}
with $\Xmlip(\Rvec)^\mathrm{T} \Xmlip(\Rvec)$ a $K\times K$ matrix. This equation corresponds to the solution of a least-squares method, as expressed in Eq.~(4) of the main text. Since the parameters $\gammavec$ appear in the right hand side of Eq.~\eqref{eq:WLS} (through the effective potential used to perform the average), a self-consistent procedure has to be employed to solve the nonlinear equation defining $\gammavec$.

\section{From the Fisher divergence to a least-squares solution}
Let us consider that the geometry of the periodic supercell is defined by a $3\times3$ matrix $\mathbf{h}$ and that $\mathbf{h}$ can change under the action of the internal or external pressure $p$. We then parametrize such changes as $\mathbf{h(\epsilonvec)}=\mathbf{h_0}(\mathbb{I}+\epsilonvec)$ with $\mathbf{h_0}$ the $3\times3$ equilibrium cell matrix, $\epsilonvec$ giving the cell strains and $\Omega (\epsilonvec)=\mathrm{det}[\mathbf{h(\epsilonvec)}]$ defining the volume (see Ref.~\cite{Cajahuaringa_JCP149,Kobayashi_JCP155_2021}). In this context, the reference and surrogate distributions become $p(\Rvec, \epsilonvec)=(\Zpart^{'} \Omega^2(\epsilonvec))^{-1} \mathrm{e}^{-\beta (V(\Rvec)+p \Omega(\epsilonvec))}$ and $\qmlip(\Rvec, \epsilonvec) = (\widetilde{\Zpart^{'}_{\gammavec}} \Omega^2(\epsilonvec))^{-1} \mathrm{e}^{-\beta (\Vmlip(\Rvec)+p \Omega(\epsilonvec))}$, with $\Omega_0=\mathrm{det}[\mathbf{h_0}]$ and the following reference and surrogate partition functions $\Zpart^{'}= \Omega_0 \int\dint\epsilonvec\Zpart\mathrm{e}^{(-\beta p \Omega(\epsilonvec))}\Omega^{-2}(\epsilonvec)$ and $\widetilde{\Zpart^{'}_{\gammavec}}= \Omega_0 \int\dint\epsilonvec\widetilde{\Zpart_{\gammavec}}\mathrm{e}^{(-\beta p \Omega(\epsilonvec))}\Omega^{-2}(\epsilonvec)$, respectively. To simplify the notation, we remove the arguments $(\Rvec, \epsilonvec)$ of the distributions, potentials, partition functions... in the following.  

The aim of this section is to obtain an equation equivalent to Eq~\eqref{eq:WLS} when the Kullback--Leibler divergence $\DKL$ is replaced by the Fisher divergence $\DF$~\cite{Yang_ArXiv_2019,Lyu_Proc_2009}. This later measures the difference between the two distributions $\qmlip$ and $p$ while including the gradient of the energy with respect to $\etavec$ (indexing the atomic positions $\Rvec$ or the cell strains $\epsilonvec$):
\begin{equation}
    \DF^{\etavec}(\qmlip \Vert p) = \iint\dint\Rvec\dint\epsilonvec \qmlip(\Rvec, \epsilonvec) \bigg\vert \nabla_{\etavec} \ln \bigg[ \frac{\qmlip(\Rvec, \epsilonvec)}{p(\Rvec, \epsilonvec)} \bigg] \bigg\vert^2 \geq 0 \mbox{\; .} \nonumber
\end{equation}
This divergence has properties similar to the Kullback--Leibler one: the smaller $\DF^{\etavec}$ is, the closer the distribution $p$ and $\qmlip$ are, with $\DF^{\etavec}(\qmlip\Vert p)=0$ meaning that the two distributions are identical. This quantity can be rewritten as:
\begin{eqnarray}
    &\DF^{\etavec}(\qmlip \Vert p) = \displaystyle\iint\dint\Rvec\dint\epsilonvec \qmlip \bigg\vert \nabla_{\etavec} \bigg[\ln \bigg(\displaystyle \frac{\Zpart}{\widetilde{\Zpart}_{\gammavec}}\bigg)-\beta \Vmlip +\beta V\bigg]  \bigg\vert^2 \nonumber \\
    &= \beta^2 \displaystyle\iint\dint\Rvec\dint\epsilonvec \qmlip \bigg\vert \nabla_{\etavec}V - \nabla_{\etavec}\Vmlip \bigg \vert^2 \mbox{\; .}
\label{eq:DF_eta}
\end{eqnarray}
Depending on whether we consider the gradients with respect to the atomic positions $\Rvec$ or to the cell strains $\epsilonvec$, we obtain the following equations:
\begin{eqnarray}
    \DF^{\Fvec}(\qmlip \Vert p)    & = & \displaystyle\beta^2\braket{ \vert \Fvec     - \widetilde{\Fvec}_{\gammavec} \vert^2 }_{\Vmlip}  \mbox{\; ,} \\ 
    \DF^{\sigmavec}(\qmlip \Vert p) & = & \displaystyle\beta^2\braket{ \vert \sigmavec - \widetilde{\sigmavec}_{\gammavec} \vert^2 }_{\Vmlip} \mbox{\; ,}
\end{eqnarray}
with $\Fvec=-\nabla_{\Rvec}V(\Rvec,\epsilonvec)$ and $\widetilde{\Fvec}_{\pmb{\gamma}}=-\nabla_{\Rvec}\Vmlip(\Rvec,\epsilonvec)$ the true and MLIP atomic forces, and $\pmb\sigma=\nabla_{\epsilonvec}V(\Rvec,\epsilonvec)$ and $\widetilde{\pmb\sigma}_{\pmb{\gamma}}=\nabla_{\epsilonvec}\Vmlip(\Rvec,\epsilonvec)$ the true and MLIP cell stresses, respectively.

As for the Kullback--Leibler divergence, we can compute the gradient of $\DF^{\etavec}$ to find out the $\gammavec$ parameters, solutions of the minimization procedure. Starting from Eq.~\eqref{eq:DF_eta} and using $\braket{\Xmlip(\Rvec)^{\mathrm{T}}}_{\Vmlip} = 0$, we obtain
\begin{eqnarray}
    \nabla_{\gammavec} \DF^{\etavec} & = & \nabla_{\gammavec} \bigg(\beta^2\braket{(\nabla_{\etavec}\Vmlip - \nabla_{\etavec}V)(\nabla_{\etavec}\Vmlip - \nabla_{\etavec}V)}_{\Vmlip} \bigg) \nonumber \\
    & = & \beta^2 \bigg( 2\braket{\nabla_{\etavec}(\nabla_{\gammavec}\Vmlip)\nabla_{\etavec}(\Vmlip - V)}_{\Vmlip} \nonumber \\
    &   & -\beta \braket{\vert\nabla_{\etavec}(\Vmlip - V)\vert^2 ( \nabla_{\gammavec}\Vmlip)}_{\Vmlip}  \bigg) \mbox{\; ,}
\end{eqnarray}
the second term on the right-hand side of the previous equation coming from the gradient of the Boltzmann weight. The latter contribution is a second-order term, much lower than the first one if the convergence towards the fixed point is reached~\cite{Yang_ArXiv_2019}. If we impose that $\nabla_{\gammavec} \DF^{\etavec}=0$ and use $\nabla_{\gammavec} \Vmlip = \Xmlip^{\mathrm{T}}$, we obtain
\begin{eqnarray}
\label{eq:LS_Fisher_tot}
    \gammavec & = & \Braket{\nabla_{\etavec}\Xmlip^\mathrm{T} \nabla_{\etavec}\Xmlip}_{\Vmlip}^{-1} \\
              &   & \times \bigg[ \Braket{ \nabla_{\etavec}\Xmlip^\mathrm{T} \nabla_{\etavec}V}_{\Vmlip} + \frac{\beta}{2} \Braket{\vert\nabla_{\etavec}(\Vmlip-V)\vert^2 \Xmlip^{\mathrm{T}}}_{\Vmlip} \bigg] \mbox{\; ,} \nonumber  
\end{eqnarray}
with $\nabla_{\etavec}\Xmlip^\mathrm{T}\nabla_{\etavec}\Xmlip$ a $K\times K$ matrix and the following convention for the gradients:
\begin{equation}
\nabla_{\etavec}\Xmlip = 
    \begin{bmatrix}
        \vdots & \vdots & \vdots & \vdots \\
        \nabla_{\etavec}\Xd_{1} & \nabla_{\etavec}\Xd_{2} & \hdots & \nabla_{\etavec}\Xd_{K} \\
        \vdots & \vdots & \vdots & \vdots \\
    \end{bmatrix}
    \mbox{\; .}
\end{equation}
The first term on the right-hand side of Eq.~\eqref{eq:LS_Fisher_tot} is the solution of a least-squares method, equivalent to Eq.~\eqref{eq:WLS}, but concerns the forces or stresses. The second term is discarded in the present version of MLACS, so we just implement: 
\begin{equation}
\label{eq:LS_Fisher}
    \gammavec = \Braket{\nabla_{\etavec}\Xmlip^\mathrm{T} \nabla_{\etavec}\Xmlip}_{\Vmlip}^{-1} \Braket{ \nabla_{\etavec}\Xmlip^\mathrm{T} \nabla_{\etavec}V}_{\Vmlip} \mbox{\; .}  
\end{equation}

\section{The MLACS procedure}

In this section, we discuss how to solve Eq.~\eqref{eq:WLS} and~\eqref{eq:LS_Fisher}. If these formulas are similar to the ones obtained with a simple linear least-squares, they show a circular dependency which requires a self-consistent procedure.

\subsection{From a continuous model to a discrete one}
Let us assume that we have a system of $3\Nat$ atoms defined by their atomic positions $\Rvec$. We next consider that we no longer have an infinity of atomic configurations but $N$ configurations equally distributed, with $\Rvec_n$ the atomic positions of the $n^{th}$ configuration.

In this context, the reference energies $V_n = V(\Rvec_n)$ of the $N$ configurations are gathered within the vector 
\begin{equation}
\mathbf{Y}_N = 
    \begin{bmatrix}
        V_1 , V_2 , \ldots , V_{N} 
    \end{bmatrix}^{\mathrm{T}} \mbox{\; ,}
\end{equation}
and all the descriptors, with $\Xd_{n,k} = \Xd_k(\Rvec_n)$ the $k^{th}$ descriptor of the $n^{th}$ configuration, are also gathered within the $N\times K$ matrix 
\begin{equation}
{\mathbf{X}_N} = 
    \begin{bmatrix}
        \Xd_{1,1}        & \hdots & \Xd_{1,K} \\
        \Xd_{2,1}        & \hdots & \Xd_{2,K} \\
        \vdots & \vdots & \vdots \\
        \Xd_{N,1}        & \hdots & \Xd_{N,K} \\
    \end{bmatrix}
    \mbox{\; .}
\end{equation}
We can then rewrite the average performed in Eq.~\eqref{eq:WLS} as a discrete summation over the $N$ configurations
\begin{equation}
    \widehat{\gammavec}_N = (\mathbf{X}_N^{\mathrm{T}}\mathbf{X}_N)^{-1}(\mathbf{X}_N^{\mathrm{T}}\mathbf{Y}_N) \mbox{\; .}
\end{equation}
When the configurations $\Rvec_n$ are independent and identically distributed according to $\ensuremath{\widetilde{q}_{\widehat{\gammavec}_N}}(\Rvec)$, it can be shown using the Law of Large Numbers that $\widehat{\gammavec}_N$ is a consistent stochastic estimator of $\gammavec$ (i.e. $\lim\limits_{N\rightarrow+\infty} \widehat{\gammavec}_N=\gammavec$).

\subsection{The weighted linear least-squares}
We consider in this section that the $N$ configurations are distributed according to a probability measure which differs from the Boltzmann--Gibbs measure $\ensuremath{\widetilde{q}_{\widehat{\gammavec}_N}}(\Rvec)$ associated with the MLIP potential with parameter $\widehat{\gamma}_N$, for instance a Boltzmann--Gibbs measure $\ensuremath{\widetilde{q}_{\gammavec^{*}}}(\Rvec)$ for a MLIP potential associated with a parameter $\gammavec^{*}$ different from $\widehat{\gammavec}_N$. In order to compute averages with respect to the target measure $\ensuremath{\widetilde{q}_{\widehat{\gammavec}_N}}(\Rvec)$, one needs to reweight each configuration by a factor $w_n$ proportional to $\widetilde{q}_{\gammavec^{*}}(R_n)/\widetilde{q}_{\widehat{\gammavec}_N}(R_n)$. Without loss of generality, we can assume that weights are normalized such that $\sum_n w_n = 1$.

In this context, the average of an observable with respect to $\ensuremath{\widetilde{q}_{\widehat{\gammavec}_N}}(\Rvec)$ 
\begin{equation}
    \braket{O}_{\ensuremath{\widetilde{V}_{\widehat{\gammavec}_N}}} = \int \dint\Rvec O(\Rvec) \ensuremath{\widetilde{q}_{\widehat{\gammavec}_N}}(\Rvec) 
\end{equation}
can be approximated by a discrete sum over the $N$ configurations:
\begin{equation}
\label{eq:average}
    \braket{O}_N = \sum_n O(\Rvec_n) w_n  \mbox{\; .}
\end{equation}
The Law of Large Numbers again shows that this approximation is consistent: $\lim\limits_{N\rightarrow+\infty} \braket{O}_N=\braket{O}_{\Vmlip}$. In this framework, the average performed in Eq.~\eqref{eq:WLS} becomes
\begin{equation}
\label{eq:WLSN}
    \widehat{\gammavec}_N = (\mathbf{X}_N^{\mathrm{T}}\mathbf{W}_N\mathbf{X}_N)^{-1}(\mathbf{X}_N^{\mathrm{T}}\mathbf{W}_N\mathbf{Y}_N) \mbox{\; ,}
\end{equation}
with $\mathbf{W}_N$ the $N\times N$ diagonal matrix of weights
\begin{equation}
\mathbf{W}_N=\mathrm{diag}\{w_1,w_2,\ldots,w_N\} \mbox{\; .}    
\end{equation}

\subsection{All data together: energies, forces and stresses}
In this section, the aim is to provide a formulation of the least-squares solution coming from the minimization of the following cost function: 
\begin{equation}
\Delta = \alpha_{\mathrm{E}} \DKL + \alpha_{\mathrm{F}} \DF^{\Fvec} + \alpha_{\mathrm{S}} \DF^{\sigmavec}
\label{eq:costfunc}
\end{equation}
with $\alpha_{\mathrm{E}}$, $\alpha_{\mathrm{F}}$ and $\alpha_{\mathrm{S}}$ parameters allowing to adjust the contributions from the energies, forces and stresses. Let us define $V_n = V(\Rvec_n,\epsilonvec_n)$, $\Fvec_n = -\nabla_{\Rvec}V_n$ (a vector with $3\Nat$ components) and $\sigmavec_n = \nabla_{\epsilonvec}V_n$ (a vector with 6 components in the Voigt notation), respectively the energy, the forces and the stresses of the $n^{th}$ configuration. Thus, we can construct the $(3\Nat + 7)N$-long label vector $\mathbf{Y}_N$ with energies, forces and stresses of the $N$ configurations:
\begin{equation}
\mathbf{Y}_N = 
    \begin{bmatrix}
        V_1 , \Fvec_1 , \sigmavec_1 , \ldots , V_{N} , \Fvec_{N} , \sigmavec_{N}
    \end{bmatrix}^{\mathrm{T}} \mbox{\; .}
\end{equation}
For the MLIP potential, we define $\Xd_{n,k} = \Xd_k(\Rvec_n,\epsilonvec_n)$, $\mathbf{f}_{n,k}= -\nabla_\Rvec\Xd_{n,k}$ and $\mathbf{s}_{n,k}= \nabla_{\epsilonvec} \Xd_{n,k}$, respectively the $k^{th}$ descriptor, its gradient with respect to the atomic positions and with respect to the cell strains for the $n^{th}$ configuration. We can then write the $(3\Nat + 7) N\times K$ feature matrix $\mathbf{X}_N$ built using the descriptors:
\begin{equation}
{\mathbf{X}_N} = 
    \begin{bmatrix}
        \Xd_{1,1}        & \hdots & \Xd_{1,K} \\
        \mathbf{f}_{1,1} & \hdots & \mathbf{f}_{1,K} \\
        \mathbf{s}_{1,1} & \hdots & \mathbf{s}_{1,K} \\
        \vdots & \vdots & \vdots \\
        \vdots & \vdots & \vdots \\
        \Xd_{N,1}        & \hdots & \Xd_{N,K} \\
        \mathbf{f}_{N,1} & \hdots & \mathbf{f}_{N,K} \\
        \mathbf{s}_{N,1} & \hdots & \mathbf{s}_{N,K} \\
    \end{bmatrix} \mbox{\; ,}
\end{equation}
and the $(3\Nat + 7) N\times (3\Nat + 7) N$ diagonal matrix $\mathbf{W}_N$ including the regularization parameters (the weights) as:
\begin{equation}
\label{eq:Weights}
\begin{split}
\mathbf{W}_N=\mathrm{diag}\{ & \alpha_{\mathrm{E}}w_1,\alphavec_{\mathrm{F}}w_1,\alphavec_{\mathrm{S}}w_1,\hdots,\\
& \alpha_{\mathrm{E}}w_N,\alphavec_{\mathrm{F}}w_N,\alphavec_{\mathrm{S}}w_N\} 
\end{split}   
\end{equation}
with $\alphavec_{\mathrm{F}}=\big[\alpha_{\mathrm{F}},...,\alpha_{\mathrm{F}}\big]$ a vector with $3\Nat$ components and $\alphavec_{\mathrm{S}}=\big[\alpha_{\mathrm{S}},...,\alpha_{\mathrm{S}}\big]$ a vector with 6 components. Using these definitions, the $\widehat{\gammavec}_N$ parameters, solutions to the least-squares problem associated with Eq.~\eqref{eq:costfunc}, are still given by \eqref{eq:WLSN}. 

\subsection{The self-consistent procedure}
\begin{figure}[h]
\includegraphics[scale = 0.12]{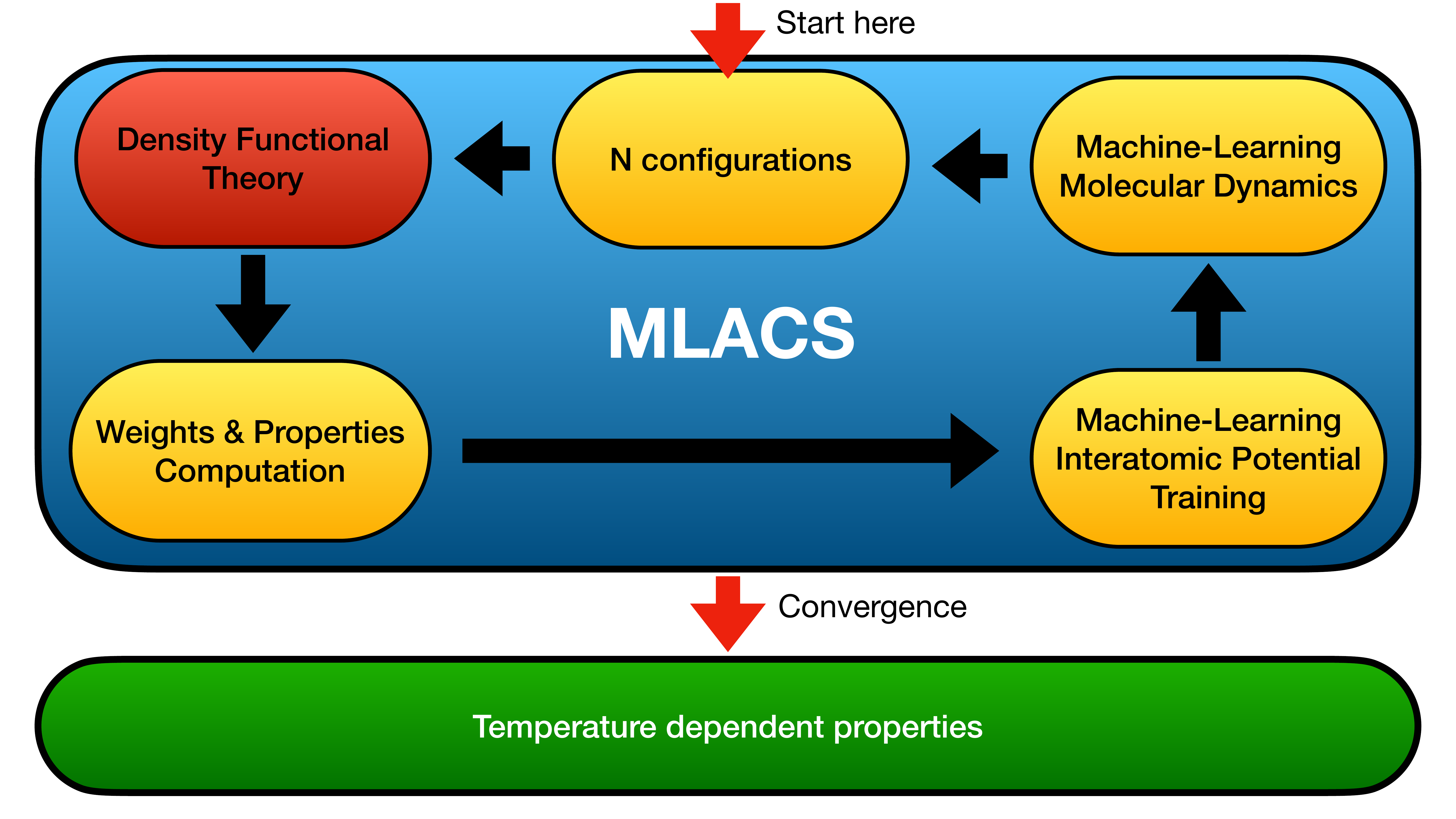}
\caption{\label{fig:MLACS} Workflow of MLACS.}
\end{figure}

Since the right-hand side of Eq.~\eqref{eq:WLSN} implicitly depends on the $\widehat{\gammavec}_N$ parameters (the reweighted configurations being distributed according to $\ensuremath{\widetilde{q}_{\widehat{\gammavec}_N}}(\Rvec)$), a self-consistent (SC) procedure is employed. The algorithm implemented in our python code is detailed in Alg.~\ref{alg:mlacs} (see also Fig.~\ref{fig:MLACS}). We stress that a crucial reweighting procedure described in section~\ref{sec:MBAR} is used within the algorithm. We consider that $N_i$ configurations are computed at step $i$, with a total number of atomic configurations $N=\sum_i N_i$. 
\begin{algorithm}[H]
\caption{\textsc{MLACS}}\label{alg:mlacs}
\begin{algorithmic}[1]
\REQUIRE Start with $N_0$ atomic configurations $\{\Rvec^{\rm (0)}_{n_{0}},\epsilonvec^{\rm (0)}_{n_{0}}\}_{n_{0}=1,...N_{0}}$ equally distributed according to $\widehat{\gammavec}_N^{(0)}$ or randomly distributed around equilibrium positions and $\mathbf{h_0}$.\\ 
\underline{Loop over SC steps}
\FOR{$i=0,1,\ldots$}
\FOR{$n_i=1,\ldots$, $N_i$}
\STATE 
\underline{DFT calculations}: \\compute $V_{n_i}^{(i)}= V(\Rvec_{n_i}^{(i)},\epsilonvec_{n_i}^{(i)})$,\\ ~~~~~~~~~~~ $\Fvec_{n_i}^{(i)}= -\nabla_{\Rvec}V_{n_i}^{(i)}$ and\\ ~~~~~~~~~~~ $\sigmavec_{n_i}^{(i)}= \nabla_{\epsilonvec}V_{n_i}^{(i)}$
\STATE \underline{MLIP calculations}:\\ compute $\Xd_{{n_i}}^{(i)}= \{\Xd_k(\Rvec_{n_i}^{(i)},\epsilonvec_{n_i}^{(i)})\}_{k=1, ...K}$,\\ ~~~~~~~~~~~ $\mathbf{f}_{{n_i}}^{(i)}= -\nabla_\Rvec\Xd_{{n_i}}^{(i)}$ and\\ ~~~~~~~~~~~ $\mathbf{s}_{{n_i}}^{(i)}= \nabla_{\epsilonvec} \Xd_{{n_i}}^{(i)}$
\ENDFOR
\STATE \underline{Reweighting (MBAR)}:\\
\FOR{$j=0,\ldots, i$}
\FOR{$n_j=1,\ldots$, N$_j$}
\STATE compute $w_{j,n_j}^{(i)}$ and build $\mathbf{W}_N^{(i)}$
\ENDFOR
\ENDFOR
\STATE \underline{Observable calculations}:\\ compute $\braket{O}^{(i)} = \sum_{n_j,j} O\big(\Rvec_{n_j}^{(j)},\epsilonvec_{n_j}^{(j)}\big) w_{j,n_j}^{(i)}$ and\\ ~~~~~~~~~~~ $\Delta\braket{O}=\vert \braket{O}^{(i)}-\braket{O}^{(i-1)} \vert$
\IF{($\Delta\braket{O}\leq$ criterion)} 
\STATE {\bf exit}
\ENDIF
\STATE \underline{MLIP fit}:\\ Build $\Xvec_N=\{\Xd_{1}^{(0)},\mathbf{f}_{1}^{(0)},\mathbf{s}_{1}^{(0)},\ldots,\Xd_{N}^{(i)},\mathbf{f}_{N}^{(i)},\mathbf{s}_{N}^{(i)}\}$ and\\ ~~~~~~~~$\Yvec_N=\{V_{1}^{(0)},\Fvec_{1}^{(0)},\sigmavec_{1}^{(0)},\ldots,V_{N}^{(i)},\Fvec_{N}^{(i)},\sigmavec_{N}^{(i)}\}$\\
\STATE Compute $\widehat{\gammavec}_N^{(i)} = (\Xvec_N^{\rm T}\Wvec_N^{(i)}\Xvec_N)^{-1} (\Xvec_N^{\rm T}\Wvec_N^{(i)}\Yvec_N)$
\STATE \underline{MLMD simulation}:\\ using $\widetilde{V}_{\widehat{\gammavec}_N^{(i)}}(\Rvec,\epsilonvec)=\sum_{k=1}^{K} \widehat{\gamma}_{N,k}^{(i)} \Xd_k(\Rvec,\epsilonvec)$ 
\STATE Extract $N_{i+1}$ atomic configurations $\{\Rvec^{(i+1)}_{n_{i+1}},\epsilonvec^{(i+1)}_{n_{i+1}}\}_{n_{i+1}=1,...N_{i+1}}$ in MLMD trajectory
\ENDFOR
\end{algorithmic}
\end{algorithm}
A criterion based on both the phonon frequencies ($\frac{\Delta\omega}{\omega} \leq 1\%$) and the pair distribution function ($\Delta g(r)\leq 0.1$) is used to stop the SC loop. 

\subsection{The MBAR reweighting\label{sec:MBAR}}
The configurations generated at every steps of the SC procedure are not distributed according to the canonical distribution of the last potential. Formally, only the data distributed according to the canonical distribution of the last $\widetilde{V}_{\widehat{\gammavec}_N^{(i)}}(\Rvec,\epsilonvec)$ can be used. In order to make use of all the configurations generated before the last step, a reweighting strategy has been implemented, based on the the Multistate Bennett Acceptance Ratio (MBAR)~\cite{Shirts2008,shirts2017reweighting,Stoltz_2010}. In this framework, it is possible to compute the weight $w_{j,n_j}^{(i)}$ of the configuration $n_j$ at a previous step $j\leq i$ for the potential built at the current step ${(i)}$ as: 
\begin{equation}
\label{eq:MBAR}
    w_{j,n_j}^{(i)}=\frac{\displaystyle{\frac{\widetilde{q}_{\widehat{\gammavec}_N^{(i)}}\big(\Rvec_{n_j}^{(j)},\epsilonvec_{n_j}^{(j)}\big)}{\widetilde{\Zpart}_{\widehat{\gammavec}_N^{(i)}}}}}{\displaystyle{\sum_{i'=0}^{i}N_{i'}\frac{\widetilde{q}_{\widehat{\gammavec}_N^{(i')}}\big(\Rvec_{n_j}^{(j)},\epsilonvec_{n_j}^{(j)}\big)}{\widetilde{\Zpart}_{\widehat{\gammavec}_N^{(i')}}}}} \mbox{\; ,}
\end{equation}
with  $\widetilde{\Zpart}_{\widehat{\gammavec}_N^{(i)}}$ determined by the nonlinear equation
\begin{equation}
    \widetilde{\Zpart}_{\widehat{\gammavec}_N^{(i)}}=\sum_{j=0}^{i}\sum_{n_j=1}^{ N_j}\frac{\widetilde{q}_{\widehat{\gammavec}_N^{(i)}}\big(\Rvec_{n_j}^{(j)},\epsilonvec_{n_j}^{(j)}\big)}{\displaystyle\sum_{i'=0}^{i}N_{i'}\frac{\widetilde{q}_{\widehat{\gammavec}_N^{(i')}}\big(\Rvec_{n_j}^{(j)},\epsilonvec_{n_j}^{(j)}\big)}{\widetilde{\Zpart}_{\widehat{\gammavec}_N^{(i')}}}} \mbox{\; .}
\end{equation}
The matrix $\Wvec_N^{(i)}$ defined by Eq.~\eqref{eq:Weights} becomes at each step ${(i)}$:
\begin{equation}
\begin{split}
\mathbf{W}_N^{(i)}=\mathrm{diag}\{&\alpha_{\mathrm{E}} w_{\rm 1,1}^{(i)},\alphavec_{\mathrm{F}} w_{\rm 1,1}^{(i)},\alphavec_{\mathrm{S}} w_{\rm 1,1}^{(i)},\hdots, \\
&\alpha_{\mathrm{E}} w_{i,N_i}^{(i)},\alphavec_{\mathrm{F}} w_{i,N_i}^{(i)},\alphavec_{\mathrm{S}} w_{i,N_i}^{(i)}\} 
\end{split}
\end{equation}
Using these weights, we are able to reuse the data computed for all the SC steps. All of them were employed to solve the least-squares method and to compute physical properties. This strategy strongly reduces the number of configurations needed to achieve the convergence.

\section{Free energy calculations\label{sec:FreeEnergy}}

Let us define the reference (DFT) free energy $\Fe$ as the sum of the surrogate (MLIP) free energy $\FeGB^0$ and a correction $\Delta \Fe$ as:
\begin{equation}
\label{eq:Fequality}
    \Fe = \FeGB^0 + \Delta \Fe \mbox{\; .}
\end{equation}
The first term on the right hand side of the previous equality can be obtained using a thermodynamic integration (TI). In this work, we use a nonequilibrium TI method for the calculation of the free energy of a solid~\cite{Freitas_2016} or liquid phase~\cite{Leite_2016,Leite_2019}. The reference systems are the Einstein crystal (with a spring constant evaluated after a short simulation) and the Uhlenbeck--Ford model (with a scaling parameter $p=50$ and a length scale parameter $\sigma=2.0$), respectively.

This TI calculation captures all the anharmonic effects included in the MLIP potential. Consequently, if the reference and surrogate distributions are almost equal at the end of the SC procedure, we obtain a good approximation of the DFT free energy $\Fe \approx \FeGB^0$. In addition, the computational cost of this calculation is really cheap, since the TI is only performed between two classical potentials. However, the DFT and MLIP distributions are sometimes slightly different, which prevents to achieve a near-DFT accuracy, with a 1 meV difference between the reference and surrogate free energies. We discuss in the following how to compute the correction $\Delta \Fe$.

According to free energy perturbation theory~\cite{Zwanzig1954}, the free energy difference in the canonical ensemble between two systems with potentials $V(\Rvec)$ and $\Vmlip(\Rvec)$ can expressed as

\begin{equation}
\begin{split}
    \Delta \Fe  &= \Fe - \FeGB^0 \\
                &= -\frac{1}{\beta} \ln \frac{\Zpart}{\widetilde{\Zpart}_{\gammavec}} \\
                &= -\frac{1}{\beta} \ln \frac{\int \dint \Rvec \mathrm{e}^{-\beta V(\Rvec)}}{\widetilde{\Zpart}_{\gammavec}} \\
                &= -\frac{1}{\beta} \ln \frac{\int \dint \Rvec \mathrm{e}^{-\beta \big(V(\Rvec)- \Vmlip(\Rvec)+ \Vmlip(\Rvec)\big)}}{\widetilde{\Zpart}_{\gammavec}} \\
                &= -\frac{1}{\beta} \ln \big\langle \mathrm{e}^{-\beta \Delta V(\Rvec)} \big\rangle_{\Vmlip}
\end{split}
\end{equation}
with $\Delta V(\Rvec)=V(\Rvec) - \Vmlip(\Rvec)$. This equation can be expanded in cumulants~\cite{Stoltz_2010} of the potential energy difference
\begin{equation}
\label{eq:cum exp}
    \Delta \Fe = \sum_{n=1}^{\infty} \frac{(-\beta)^{n-1} \kappa_n}{n!} \mbox{\; ,}
\end{equation}
where $\kappa_n$ is the $n$-th order cumulant of the potential energy differences.
Usually, the expansion~\eqref{eq:cum exp} is truncated at 2nd order, using only the first and second order cumulants of the potential energy difference: 
\begin{equation}
\begin{split}
    \kappa_1 &= \braket{\Delta V(\Rvec)}_{\Vmlip}  \mbox{\; ,}\\
    \kappa_2 &= \braket{\Delta V(\Rvec)^2}_{\Vmlip} - \braket{\Delta V(\Rvec)}_{\Vmlip}^2  \mbox{\; .}
\end{split}
\end{equation}
In this context, only an approximate free energy difference is obtained, except for systems where the potential energy difference follows a Gaussian distribution. Using this cumulant expansion, the free energy difference becomes 
\begin{equation}
    \Delta\Fe \approx \braket{\Delta V(\Rvec)}_{\Vmlip} -\frac{\beta}{2} \bigg(\braket{\Delta V(\Rvec)^2}_{\Vmlip} - \braket{\Delta V(\Rvec)}_{\Vmlip}^2\bigg)  \mbox{\; .} \nonumber
\end{equation}
Since the minimization of the Gibbs--Bogoliubov free energy is equivalent to the Kullback--Leibler Divergence minimization (see section~\ref{sec:KLD}), one can relate these terms to the information in $p(\Rvec)$ not contained in $\widetilde{q}_{\gammavec}(\Rvec)$. Consequently, this second-order cumulant free energy difference gives not only a correction to the Gibbs--Bogoliubov free energy, but also measures the accuracy of MLACS. The smaller $\Delta\Fe$ is, the better the sampling done by MLACS is. In addition, since the only data needed to evaluate this correction are the energies coming from the atomic configurations already computed by the DFT and MLIP potentials, this quantity can be estimated during a MLACS simulation at no extra cost (see Table 1 in the main paper, for systems simulated with classical potentials).

\section{Implementation of MLACS}
An implementation of MLACS, which generates configurations minimizing the cost function (see Eq.~\eqref{eq:costfunc}), has been performed in a python code containing 20 files and 5000 lines. This one is built around the  Atomic Simulation Environment (ASE) package~\cite{HjorthLarsen2017}, which allows to create supercells, launch {\it ab initio} calculations, manage I/O and perform MD simulations.

MLACS needs initial configurations as a starting point. Two choices are possible: to generate them using a MLIP potential coming from a previous calculation or to create initial random configurations. Once the first MLIP potential is created, the MLMD simulations use the Langevin thermostat implemented in the ASE package. To ensure that the configurations sample the equilibrium canonical ensemble, we perform a long trajectory (around 10 ps) and discard the beginning of this trajectory (1 or 2 ps). The MLIP potential being computationally inexpensive to evaluate compared to {\it ab initio} calculations, this step is generally fast.

Concerning the MLIP potentials, any of them can be used. For the sake of simplicity and robustness, we use the Spectral Neighbor Analysis Potential (SNAP)~\cite{Thompson2015,Wood2018,Cusentino2020} as implemented in the Large-scale Atomic/Molecular Massively Parallel Simulator (LAMMPS) package~\cite{Plimpton1995}. Its descriptors are based on the projection of the atomic environment onto a basis of hyperspherical harmonics in four dimensions. For all the MLACS simulations shown in the present work, the SNAP potential uses a 2$J_{\rm max}$ = 8 parameter and a cutoff radius of the potential adjusted at the first iteration. When two types of atoms are present in the simulation box, the weight of the atom type $i$ in the SO(4) descriptor is $h_i = Z_i/\sum_j Z_j$, with $Z_i$ the atomic number of atom type $i$.

Two additional python packages are used in the ASE environment of MLACS. The $\widehat{\gammavec}_N$ parameters (see Eq.~\eqref{eq:WLSN}) are obtained using a singular value decomposition, as implemented in the package \textsc{Numpy}. The reweighting using MBAR (see Eq.~\eqref{eq:MBAR}) has been performed by means of the \textsc{Pymbar} package~\cite{Shirts2008,shirts2017reweighting}.

The {\it ab initio} calculations were performed using the ABINIT package~\cite{Gonze2020}. This part being the most time consuming, we adopted a parallelization strategy. All the configurations extracted from the MLMD trajectory (see step 19 of Alg.~\ref{alg:mlacs}) can be computed in parallel (see step 3 of Alg.~\ref{alg:mlacs}). For instance, if only 100 configurations are needed to converge and if 20 configurations are extracted at each SC step and launched in parallel at the same time, the human waiting time is then strongly reduced to the equivalent of 5 {\it ab initio} calculations at all. If one considers that an AIMD trajectory requires around 5000 time steps, the acceleration in this case is exactly of 3 orders of magnitude. 

The phonon spectra were computed using TDEP~\cite{Hellman_PRB84_2011,Hellman2013} as implemented in the ABINIT package~\cite{Bottin_CPC_2020}. Using this method proposed by Hellman and coworkers, the renormalized phonon spectra of systems can be obtained as a function of the temperature. In this framework, the explicit (intrinsic) temperature effects are taken into account in an effective way using atomic displacements and forces coming from {\it ab initio} calculations. The number of configurations generated using MLACS being low, we implemented an improved estimator of the pair distribution function (PDF) which uses the force sampling~\cite{Rotenberg_JCP_2020}.

\section{Classical simulations with MLACS, MD and EHCS}
The aim of this section, and the next one, is to empirically demonstrate that the MLACS method is reliable. We show that MLACS \emph{converges} below a fixed tolerance criterion, is \emph{accurate} with data in excellent agreement with the reference (the MD simulation) and is \emph{efficient} with a computational cost divided by one or two orders of magnitude compared to plain AIMD simulations.

In this section and for all the examples of the following section, the initial atomic configurations are generated using random displacements around equilibrium positions. The molecular dynamic simulations are performed within the NVT ensemble. The $\alpha_{\rm E}$ and $\alpha_{\rm F}$ parameters are equal to 1 whereas $\alpha_{\rm S}=0$. The convergence is achieved when the variation of the phonon frequencies (over all the Brillouin zone) is lower than $\Delta\omega\leq0.5$~meV.

In addition to the phonon spectra, the use of TDEP makes it possible to compute other thermodynamic or elastic quantities~\cite{Bottin_CPC_2020}. In the following, we also report the free energy of the effective harmonic system $\FeTDEP^0$, the TDEP free energy $\FeTDEP = \FeTDEP^0 + \braket{V(\Rvec) - \widetilde{V}_{\mathrm{TDEP}}(\Rvec)}$, the TDEP entropy $\mathcal{S}_{\mathrm{TDEP}}$ and the elastic constants $C_{ij}$ (see equations (26), (29) and (33) in Ref.~\cite{Bottin_CPC_2020}). Using TDEP, the free energy of {\it ab initio} simulations can be computed without performing hundreds of thousands AIMD time steps, as required by thermodynamic integration (which would be out of reach for various systems shown below). For systems using classical potentials, we also perform a more rigorous comparison between reference and surrogate free energies (see Eq.~\eqref{eq:GB}) using thermodynamic integration and cumulant expansion (see section~\ref{sec:FreeEnergy} and Table I of the main text). 

We first use classical potentials. The computational time being very low for them we can thus perform long MD trajectory simulations. This allows to obtain very well converged reference results and assess the convergence and accuracy of the MLACS method.

\subsubsection{\label{sec:SiTersoff}Silicon with a Tersoff potential at $\mathrm{T}=1500~\mathrm{K}$}

Here we consider the diamond phase of Silicon at high temperature ($\mathrm{T}=1500~\mathrm{K}$) and equilibrium volume ($\mathrm{V}=20.13~\mathrm{\AA/at}$). For the Tersoff potential~\cite{Tersoff_PRL56_1986}, this temperature is close to the melting temperature, so the system is strongly anharmonic, which makes it a non-trivial test case. We use a supercell built as $(3\times3\times3)$ the conventional cell, with 216 atoms.

\begin{figure}
    \centering
    \includegraphics[width=\linewidth]{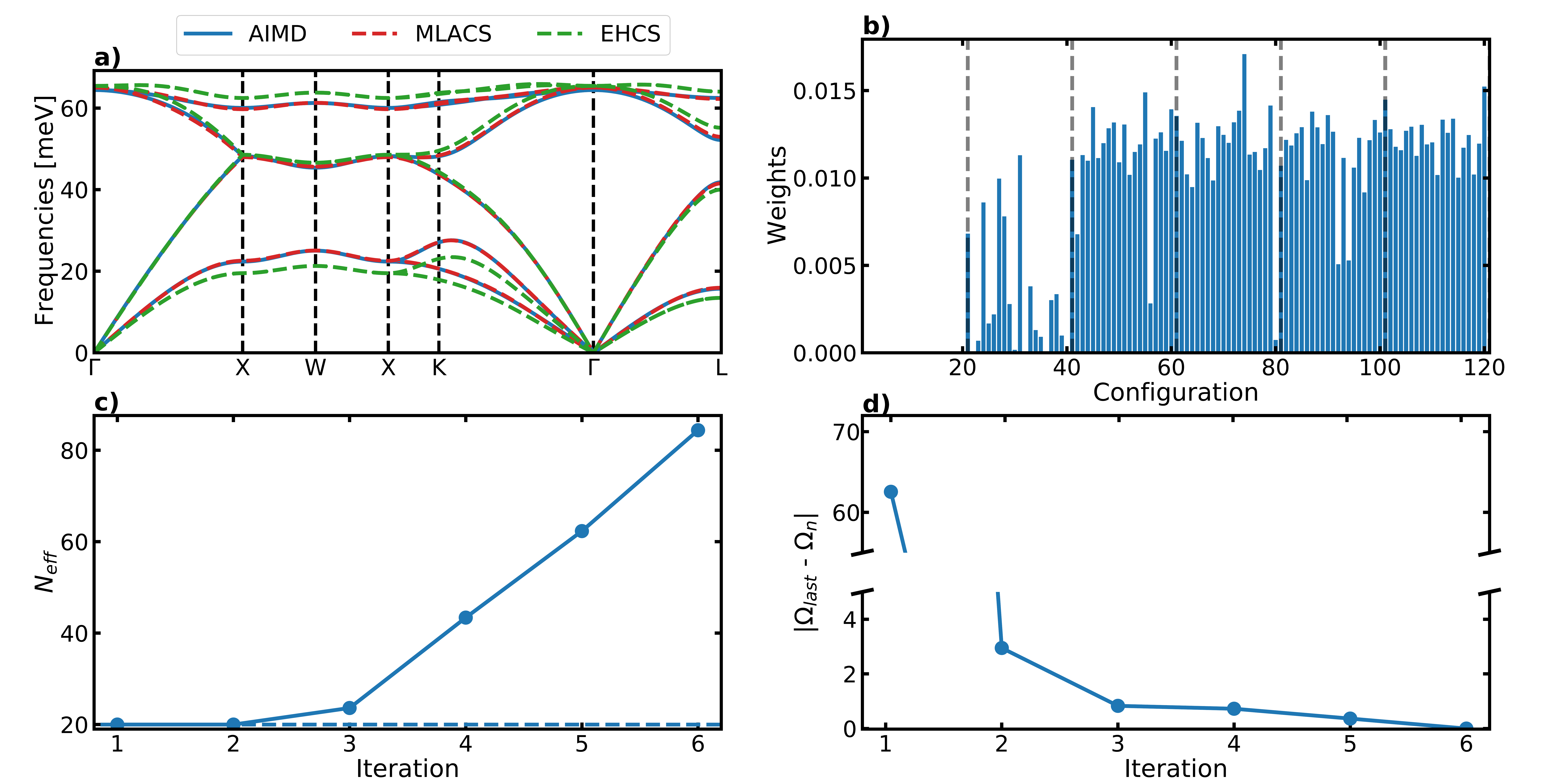}
    \caption{Results for the Tersoff potential of Silicon at $\mathrm{T}=1500~\mathrm{K}$.
             a) Comparison of the phonon spectra computed using TDEP for MD (blue), MLACS (red) and EHCS (green) simulations.
             b) Weight of each configuration for the MLACS simulation. The vertical hashed lines split the configurations coming from different steps.
             c) Effective number of configurations a each step of the MLACS simulation. The horizontal hashed line stands for the number of configurations per steps (20).
             d) Convergence of the MLACS phonon spectrum as a function of the iteration step.
         }
        \label{fig:Compare Si Tersoff}
\end{figure}

The MLACS, MD and EHCS renormalized phonon spectra are displayed on Fig.~\ref{fig:Compare Si Tersoff} a). The agreement between MLACS and MD is excellent. At odds, EHCS overestimates (under-estimates) the frequencies of the optic (acoustic) branches with respect to AIMD.

The weight of each configuration used by MLACS is given in Fig.~\ref{fig:Compare Si Tersoff} b). This picture illustrates that the reweighting process works very well. The first configurations have very small weights because they are far from the equilibrium canonical distribution, whereas the last ones equally contribute. This effect can be measured at each iteration $i$ using the effective number of configurations: 
\begin{equation}
    N_{\mathrm{eff}}^{(i)} = \frac{\big(\sum_{j,n_j} w_{j,n_j}^{(i)}\big)^2}{\sum_{j,n_j} (w_{j,n_j}^{(i)})^2}
\end{equation}
The evolution of this quantity as a function of the step number is shown in Fig.~\ref{fig:Compare Si Tersoff} c). When the MLIP potential is not well fitted, $N_{\mathrm{eff}}$ remains equal to 20 whereas $N_{\mathrm{eff}}$ increases linearly close to convergence of the SC loop.

This trend leads to a convergence of the phonon frequencies below the tolerance criterion 0.5 meV, as shown in Fig.~\ref{fig:Compare Si Tersoff} d). Without any reweighting, the accuracy would be proportional to $\frac{1}{\sqrt{N_{\mathrm{last}}}}$ (with $N_{\mathrm{last}}$ the number of atomic configurations for the last step) which would prevent any convergence below a fixed tolerance criterion. Here, the accuracy is proportional to $\frac{1}{\sqrt{N_{\mathrm{eff}}}}$ which can decrease by increasing the number of SC steps or the number of configurations for each SC step. This feature has been highlighted for EHCS simulations~\cite{Tidholm2020}, but for lower temperature conditions, so with a lower variance.

\begin{figure}
    \centering
    \includegraphics[width=1.0\linewidth]{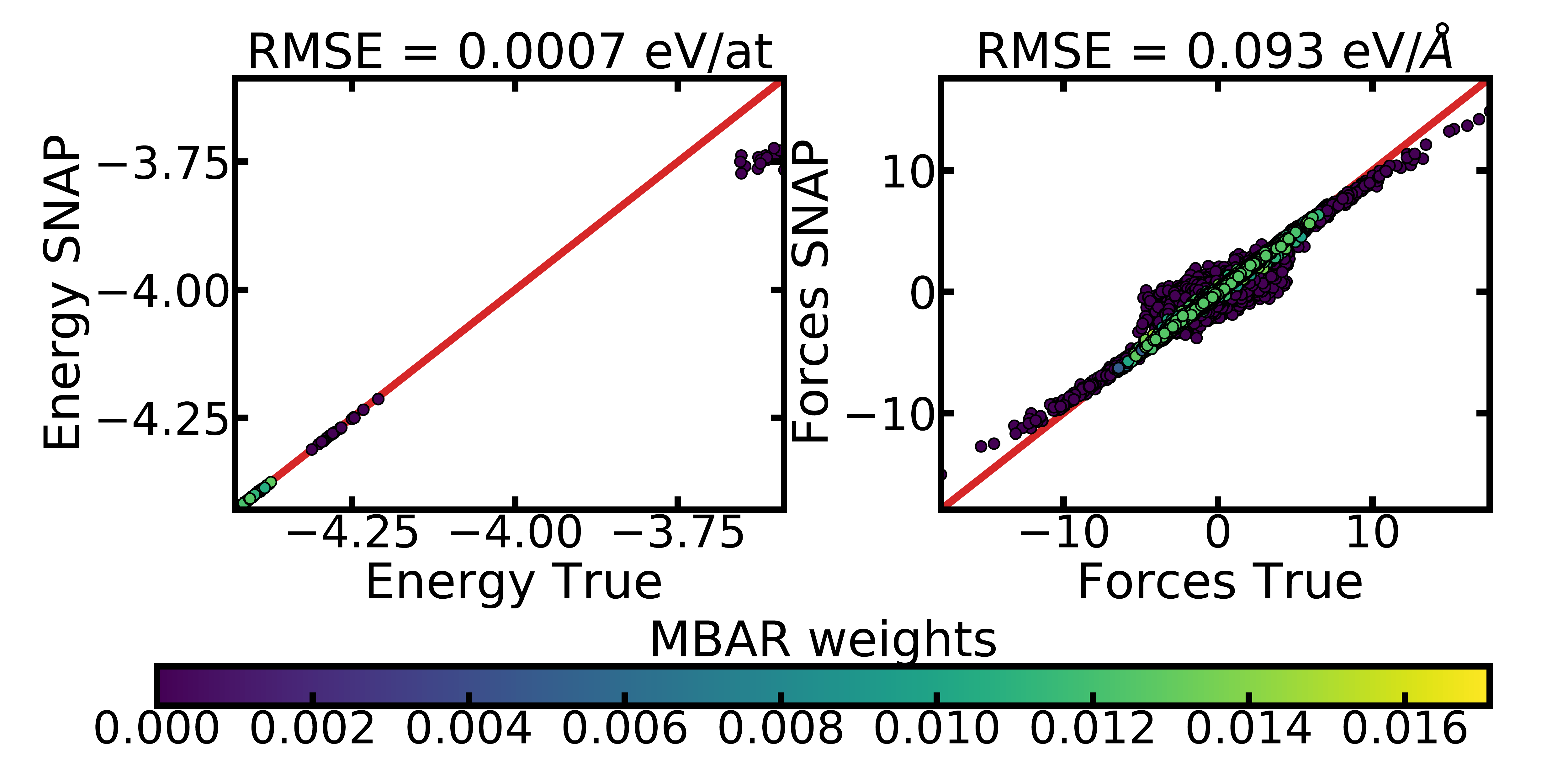}
    \caption{Correlations between the MLIP and reference data, for energies (left) and forces (right), for Silicon with the Tersoff potential at $\mathrm{T}=1500~\mathrm{K}$. The color points out the final weight of each configuration and the first bisector corresponds to linear correlations.}
    \label{fig:Fit Silicium Tersoff}
\end{figure}

In Fig.~\ref{fig:Fit Silicium Tersoff} we show the correlations between the MLIP and Tersoff energies (left) and forces (right). The color points out the final weight of each configuration. Only the points close to the first bisector contribute to the construction of the MLIP potential, the equilibrium canonical distribution and averages. These results highlight the strong resilience of MLACS with respect to extrapolation. This good description can be measured by the Root Mean Square Error (RMSE) at iteration $i$ computed for the energy as:
\begin{equation}
	\mathrm{RMSE}^{(i)} = \sqrt{\sum_{j,n_j} w_{j,n_j}^{(i)} \bigg(V(\Rvec_{n_j}^{(j)}) - \widetilde{V}_{\gammavec^{(i)}}(\Rvec_{n_j}^{(j)})\bigg)^2} \mbox{\; .} \nonumber
\end{equation}
At convergence of the SC loop, we obtain a RMSE equal to 1.2 meV/at for the energy and 183 meV/\AA~for the forces. The fit seems really good, even if a better accuracy would be obtained with some more complex MLIP (such as Moment Tensor Potentials (MTP), Atomic Cluster Expansion (ACE) descriptors or quadratic SNAP models). 

\begin{table*}
\caption{\label{tab:Compare Si Tersoff} Results for the Tersoff potential of Silicon at $\mathrm{T}=1500~\mathrm{K}$ for MD, EHCS and MLACS simulations. $\FeTDEP^0$ is the free energy of the effective harmonic system, $\FeTDEP$ is the TDEP free energy, $\mathcal{S}_{\mathrm{TDEP}}$ is the TDEP entropy and $C_{ij}$ are the elastic constants. The number of configurations $N_{\mathrm{confs}}$ takes into account all the configurations, even the initialization (MLACS and EHCS) and thermalization (AIMD) steps.}
\begin{tabular}{c c c c c c c c }
\hline \hline\\ [-0.3cm]
& $\FeTDEP^0$ (eV/at) & $\FeTDEP$ (eV/at) & $\mathcal{S}_{\mathrm{TDEP}}$ ($\mathrm{k_B}$/at) & $C_{11}$ (GPa) & $C_{12}$ (GPa)  & $C_{44}$ (GPa) & N$_{\mathrm{confs}}$\\
\hline
        AIMD  & -0.487 & -5.083 & 6.800 & 126 & 81 & 96  & 20000\\
        EHCS             & -0.499 & -5.052 & 6.894 & 122 & 94 & 106 & 320 \\
        MLACS & -0.487 & -5.083 & 6.796 & 123 & 83 & 96  & 140 \\
\hline \hline
\end{tabular}
\end{table*}

In addition to the phonon spectra (see Fig.~\ref{fig:Compare Si Tersoff}), we also show the TDEP free energies, the TDEP entropies and elastic constants in Tab.~\ref{tab:Compare Si Tersoff}. As expected, the agreement between MLACS and MD simulations is excellent, with an error lower than 1 meV/at, despite the low number of configurations required (140 rather than 20,000). For the EHCS simulations, the result is less good with an error equal to 12 meV/at for $\FeTDEP^0$ and 31 meV/at for $\FeTDEP$. The conclusion of this section is that MLACS can achieve a DFT accuracy for anharmonic systems.

\subsubsection{Al$_{0.5}$Cu$_{0.5}$ alloy with an ADP potential at $\mathrm{T}=600~\mathrm{K}$}

\begin{figure}
    \centering
    \includegraphics[width=\linewidth]{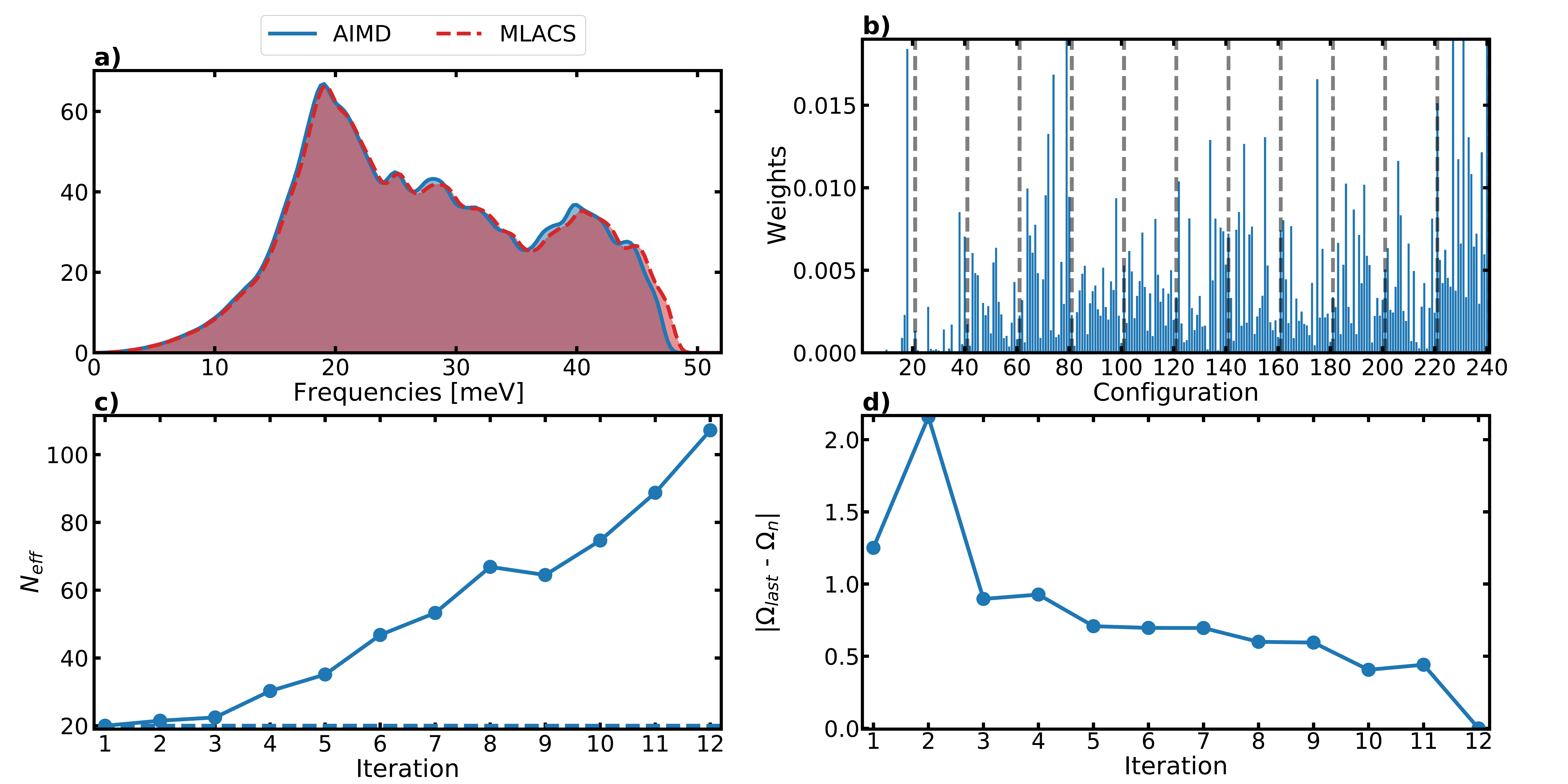}
    \caption{Same caption as Fig.~\ref{fig:Compare Si Tersoff} for the Al$_{0.5}$Cu$_{0.5}$ alloy at $\mathrm{T}=600~\mathrm{K}$. a) now displays the DOS computed using MD-TDEP (blue) and MLACS-TDEP (red).}
    \label{fig:Compare AlCu}
\end{figure}

The second example is the Al$_{0.5}$Cu$_{0.5}$ alloy in its fcc phase, at $\mathrm{T}=600~\mathrm{K}$ and for a volume $\mathrm{V}=13.078~\mathrm{\AA/at}$. This system is simulated using an Angular-Dependent Potential (ADP)~\cite{Mishin_AM53_2005} with 108 atoms in the supercell. In this case, we do not show the phonon spectrum (which is more complex due to phonon scattering) but only plot the phonon density of states (DOS) in Fig.~\ref{fig:Compare AlCu} a). As previously, the agreement between MLACS and MD simulations is excellent. This result comes from the efficiency of the reweighting (see Fig.~\ref{fig:Compare AlCu} b)) since almost all the configurations contribute to the average. Finally, we obtain only slight differences between the thermodynamic and elastic properties of the ADP and MLIP potentials (see Tab.~\ref{tab:Compare AlCu}), which opens the way to using MLACS for the simulation of alloys.

\begin{figure}
    \centering
    \includegraphics[width=1.0\linewidth]{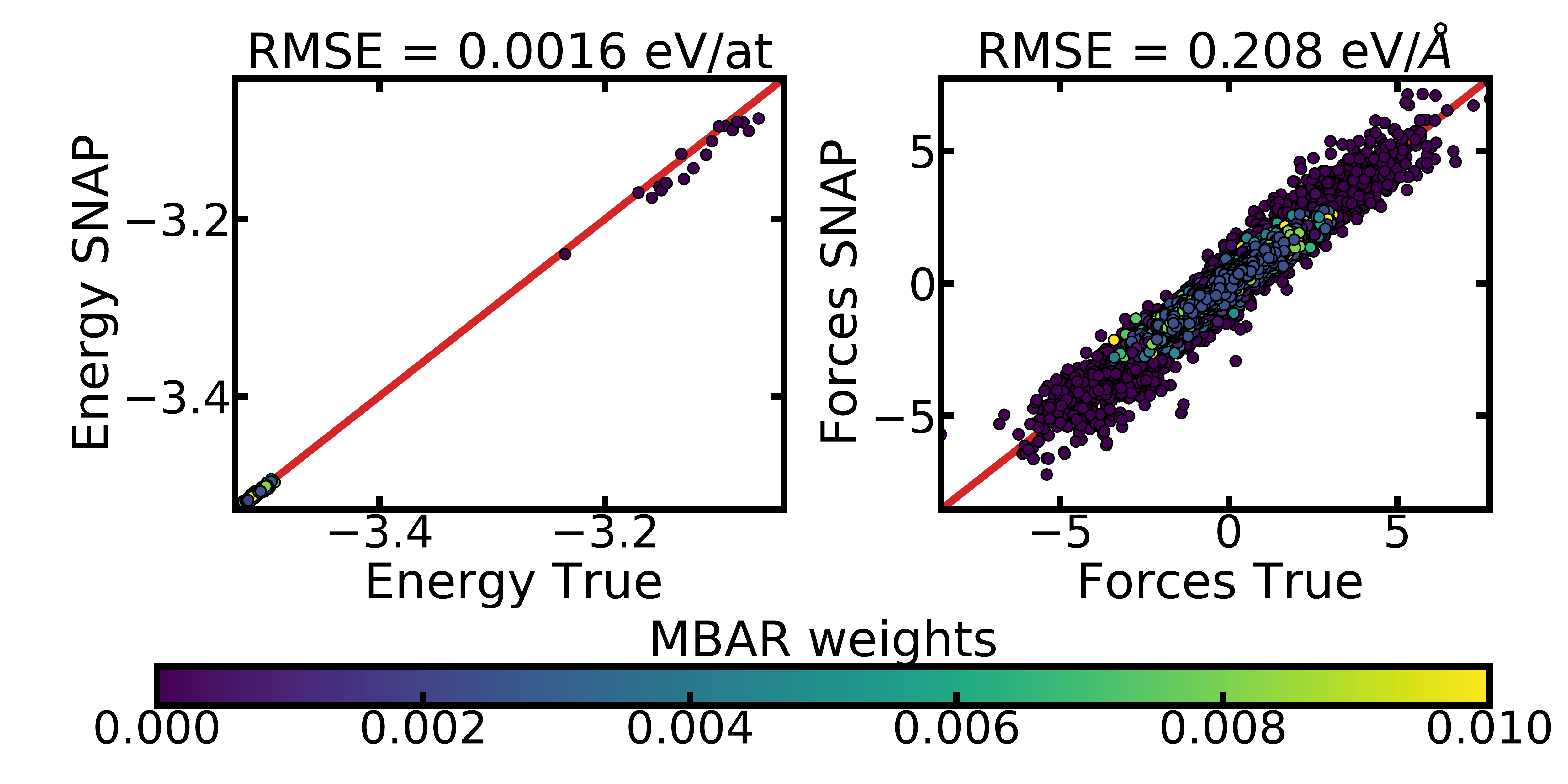}
    \caption{Same caption as Fig.~\ref{fig:Fit Silicium Tersoff} for the Al$_{0.5}$Cu$_{0.5}$ alloy at $\mathrm{T}=600~\mathrm{K}$.}
    \label{fig:Fit AlCu}
\end{figure}

\begin{table*}
\centering
    \caption{\label{tab:Compare AlCu} Same caption as Tab.~\ref{tab:Compare Si Tersoff}  for the Al$_{0.5}$Cu$_{0.5}$ alloy at $\mathrm{T}=600~\mathrm{K}$.}
\begin{tabular}{c c c c c c c c }
\hline \hline\\ [-0.3cm]
& $\FeTDEP^0$ (eV/at) & $\FeTDEP$ (eV/at) & $\mathcal{S}_{\mathrm{TDEP}}$ ($\mathrm{k_B}$/at) & $C_{11}$ (GPa) & $C_{12}$ (GPa)  & $C_{44}$ (GPa) & N$_{\mathrm{confs}}$\\
\hline
        MD    & -0.104 & -3.731 & 5.081 & 90 & 32 & 31 & 20000 \\
        MLACS & -0.103 & -3.724 & 5.074 & 86 & 36 & 31 & 260   \\
\hline \hline
\end{tabular}
\end{table*}

\subsubsection{Uranium liquid with a MEAM potential at $\mathrm{T}=2500~\mathrm{K}$}
The last example based on classical simulations is the Uranium liquid at $\mathrm{T}=2500~\mathrm{K}$ with a volume $\mathrm{V}=20.05~\mathrm{\AA/at}$. This system is simulated using a Modified Embedded-Atom Method (MEAM) potential~\cite{Baskes_PRB46_1992} with 128 atoms in the supercell. EHCS cannot be performed on a liquid, so we will only compare MLACS to the MD simulations. Since we cannot extract the phonon spectrum, the tolerance criterion used to check the convergence will be the PDF. The starting point of all the simulations is the perfect bcc crystal, which is far from being typical under the equilibrium distribution to sample.

\begin{figure}
    \centering
    \includegraphics[width=\linewidth]{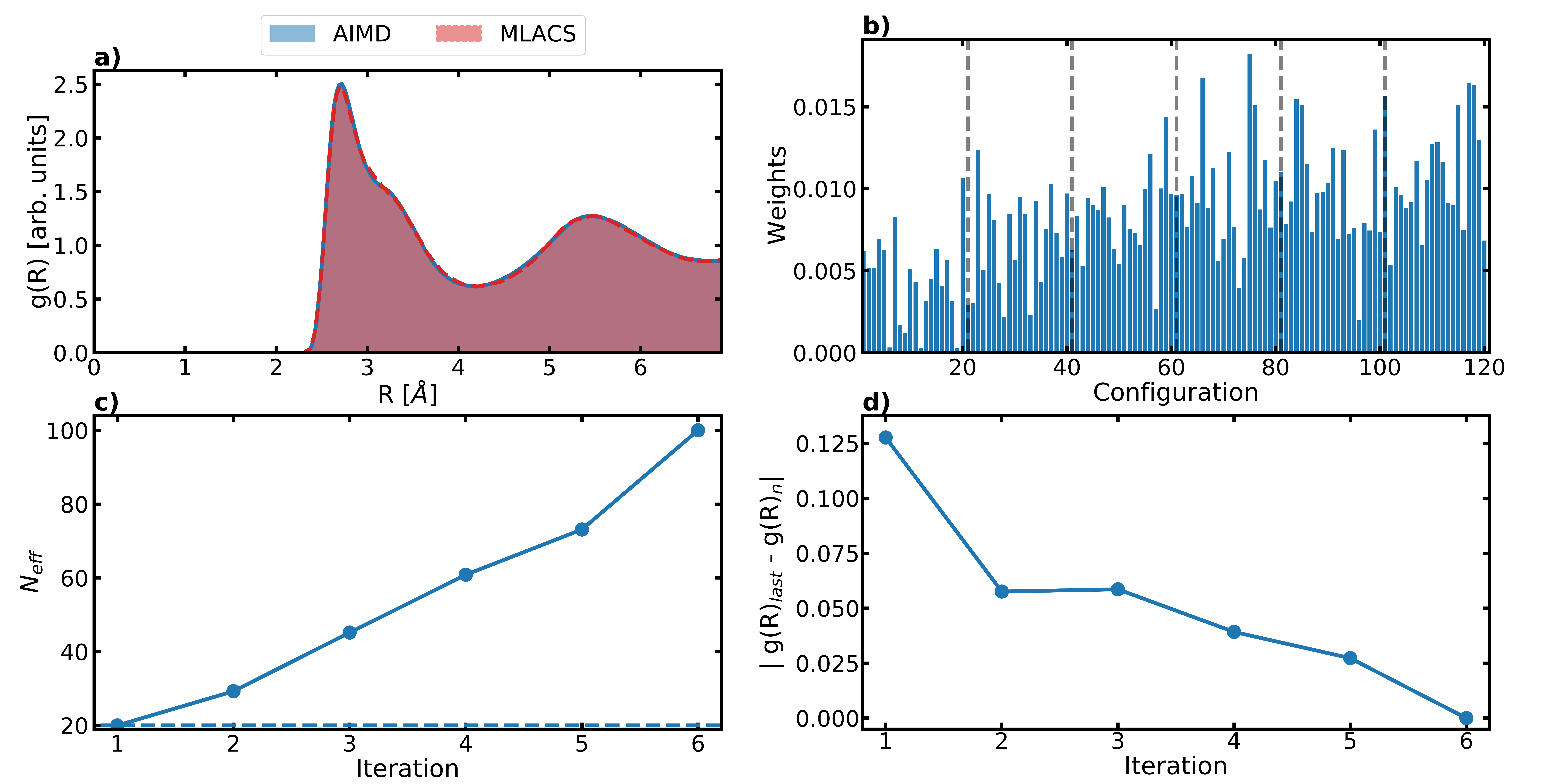}
    \caption{Same caption as Fig.~\ref{fig:Compare Si Tersoff} for the MEAM potential of Uranium at $\mathrm{T}=2500~\mathrm{K}$. a) now displays the PDF $g(r)$ computed using MD (blue) and MLACS (red).}
         \label{fig:Compare U MEAM}
\end{figure}

\begin{figure}
    \centering
    \includegraphics[width=1.0\linewidth]{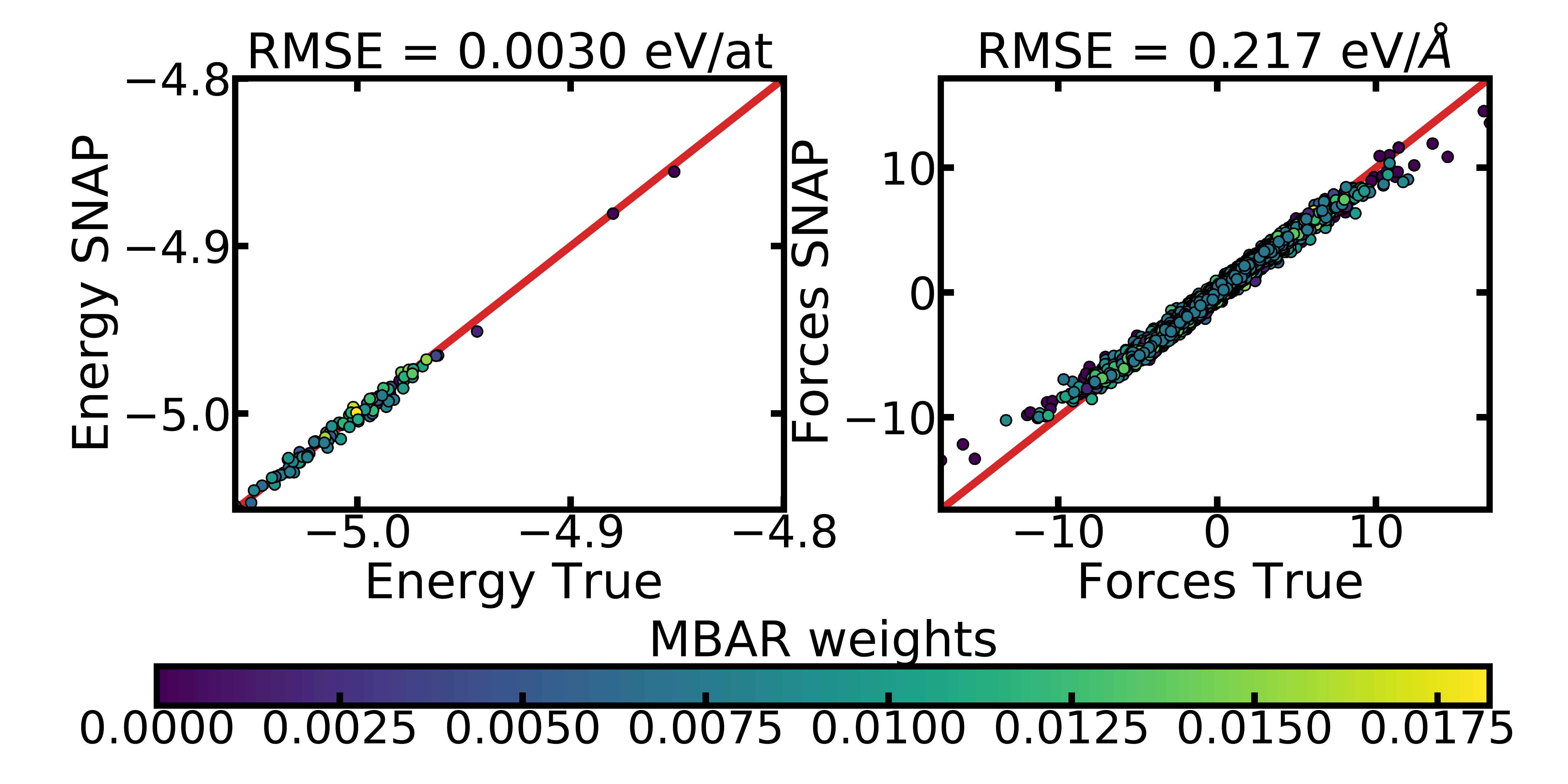}
    \caption{Same caption as Fig.~\ref{fig:Fit Silicium Tersoff} for the MEAM potential of Uranium at $\mathrm{T}=2500~\mathrm{K}$.}
    \label{fig:Fit Uranium MEAM}
\end{figure}

The results displayed in Fig.~\ref{fig:Compare U MEAM} suggest that MLACS can be applied to liquid phases. An excellent agreement is obtained between the PDF of MLACS and MD, despite the very low number of configurations performed by MLACS (140) with respect to MD (20,000). This can also be seen in Fig~\ref{fig:Fit Uranium MEAM} displaying the correlations between the MLIP and MEAM potentials. At last, even if the energy variance is high for a liquid at this temperature, the difference between average energies only equals 6 meV/at (see Tab.~\ref{tab:Compare U MEAM}). This result, associated with the good description of the bulk free energy (see section\ref{sec:SiTersoff}), opens the way to an accurate and fast description of the melting curve. 

\begin{table}
\centering
\caption{\label{tab:Compare U MEAM} Average energy and number of configurations for the MD and MLACS simulations. }
\begin{center}
\begin{tabular}{c c c }
\hline \hline
& $\braket{V(\Rvec)}$ (eV/at) & N$_{\mathrm{confs}}$\\
\hline
        MD    &  -5.009 &  20000\\
        MLACS &  -5.003 &  140 \\
\hline \hline
\end{tabular}
\end{center}
\end{table}

Through these three examples we demonstrate that MLACS is reliable: $(i)$ the method (and in particular the SC loop) \emph{converges} below a tolerance criterion based on phonon frequencies or PDF, $(ii)$ the method is \emph{accurate}, with a perfect agreement for an anharmonic crystal (Silicon) and a very good agreement for an alloy or a liquid, $(iii)$ the method is \emph{efficient} with a number of configurations which is strongly reduced with respect to MD calculations, and $(iv)$ last but not least, this method is able to sample the equilibrium canonical distribution of an alloy or a liquid, for which the EHCS method fails. However, the aim is to accelerate AIMD calculations and these examples cannot highlight the ability of MLACS to lead to a strong reduction of the computational time. This point is demonstrated in the main text and in the five following examples.

\section{{\it Ab initio} simulations with MLACS, AIMD and EHCS}
In this section, all the {\it ab initio} calculations are performed using the ABINIT code~\cite{Gonze2020} but we emphasize that MLACS can of course be used with any DFT code.

\subsubsection{Silicon at $\mathrm{T}=900~\mathrm{K}$}
As a first {\it ab initio} example, we consider the diamond phase of Silicon at $\mathrm{T}=900~\mathrm{K}$ and equilibrium volume $\mathrm{V}=20.13~\mathrm{\AA/at}$, with 216 atoms in the supercell. It has been shown using EHCS that anharmonicity appears at this temperature~\cite{Kim2018,Kim2020}, even if Silicon is more harmonic at lower temperatures~\cite{Knoop2020}. We use a norm-conserving pseudopotential, with a cutoff energy equal to 15 Ha, a $(2\times2\times2)$ \textbf{k}-point mesh and an exchange and correlation energy treated with the local density approximation (LDA) using the Teter parametrization. The time step used for the molecular dynamic simulations is $\Delta t \approx 2.4~\mathrm{fs}$.

\begin{figure}
    \centering
    \includegraphics[width=\linewidth]{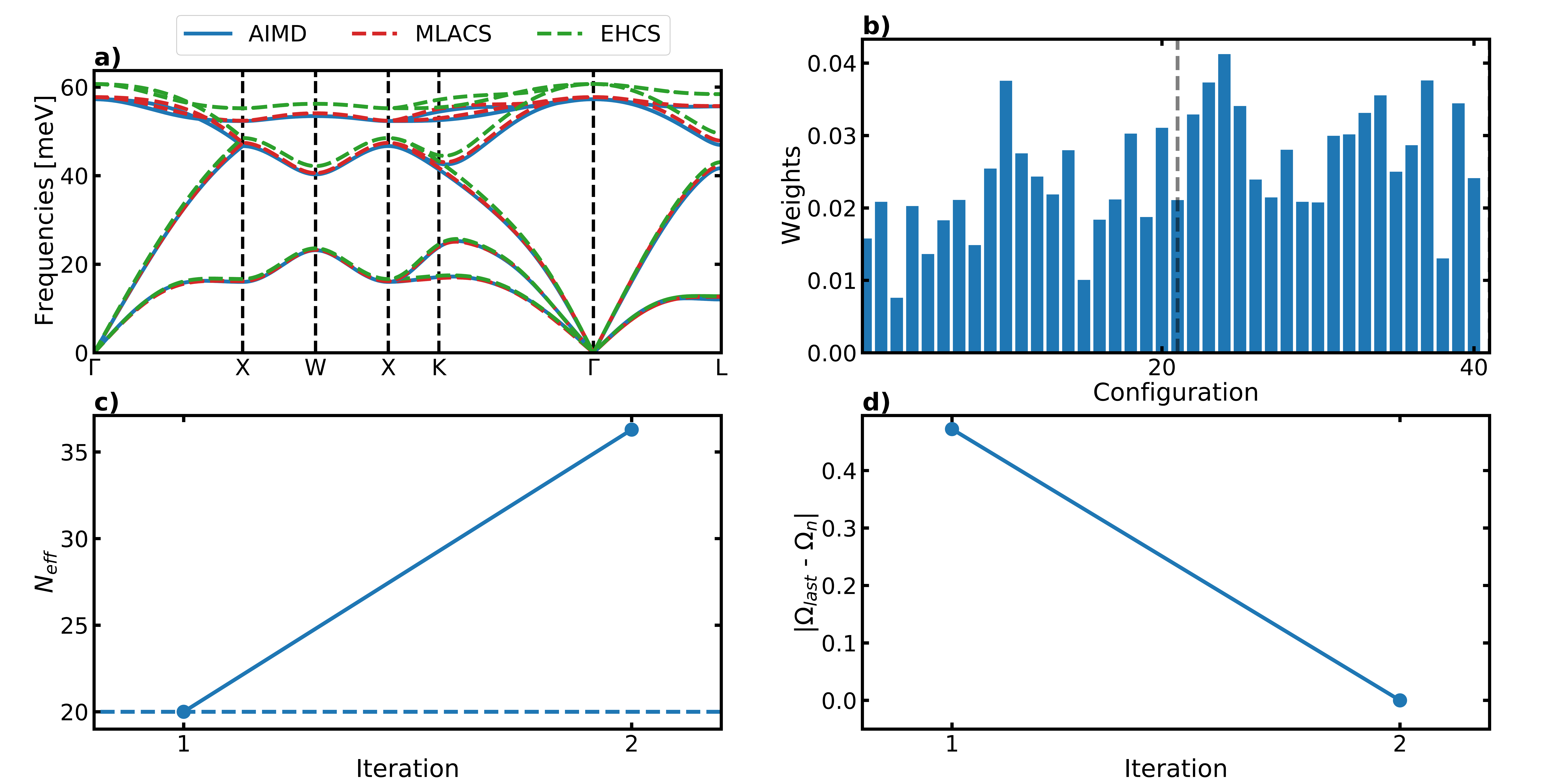}
    \caption{Same caption as Fig.~\ref{fig:Compare Si Tersoff} for Silicon at $\mathrm{T}=900~\mathrm{K}$.}
    \label{fig:Compare Si}
\end{figure}

\begin{figure}
    \centering
    \includegraphics[width=1.0\linewidth]{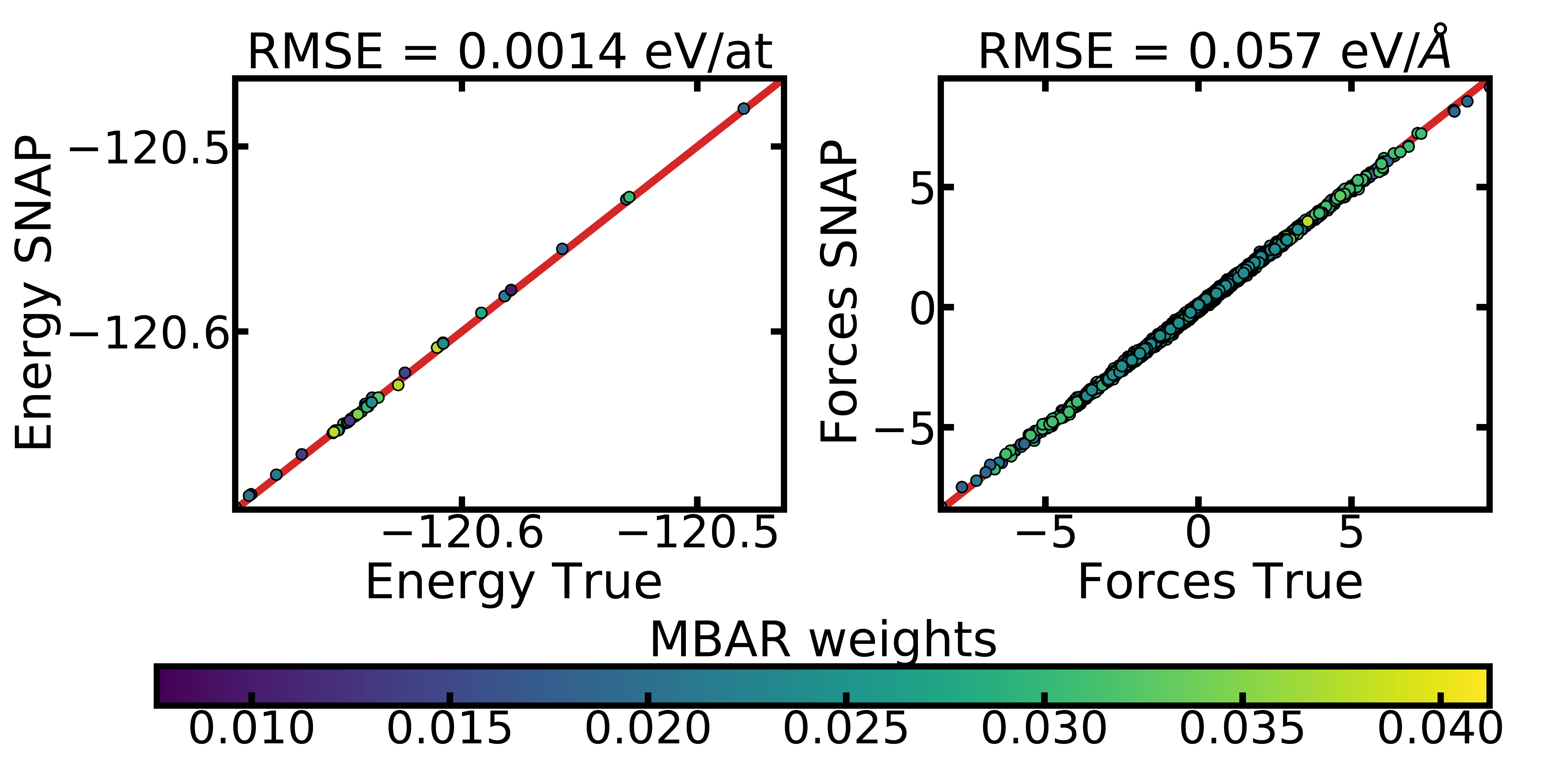}
    \caption{Same caption as Fig.~\ref{fig:Fit Silicium Tersoff} for Silicon at $\mathrm{T}=900~\mathrm{K}$.}
    \label{fig:Fit Silicium}
\end{figure}

All the results are shown in Fig.~\ref{fig:Compare Si}. The MLACS procedure stops the SC loop at the second step (after the initialization step) since $\Delta\omega\leq0.5$~meV. Despite the early stop, the MLACS results are in excellent agreement with the ones obtained from AIMD simulations, but with 50 times fewer configurations. In contrast, the EHCS simulations overestimate the optical branches by several meV.  
In Fig.~\ref{fig:Fit Silicium} we show that the reference system (DFT) is very well reproduced by the model (MLIP). With respect to other systems, the low anharmonic behavior of Silicon at $\mathrm{T}=900~\mathrm{K}$ explains the fast convergence of MLACS.

The thermodynamic and elastic properties extracted using TDEP are summarized in Tab.~\ref{tab:Compare Si}. For all of them, MLACS gives data in very good agreement with respect to AIMD, whereas EHCS simulations show some discrepancies (5 meV for the free energy, for example). Obtaining one configuration takes about 3 minutes over 2000 processors, so the MLACS results are available in 3 hours, whereas the whole AIMD trajectory needs about one week.  
\begin{table*}
\centering
    \caption{\label{tab:Compare Si} Same caption as Tab.~\ref{tab:Compare Si Tersoff} for Silicon at $\mathrm{T}=900~\mathrm{K}$.}
\begin{tabular}{c c c c c c c c }
\hline \hline\\ [-0.3cm]
& $\FeTDEP^0$ (eV/at) & $\FeTDEP$ (eV/at) & $\mathcal{S}_{\mathrm{TDEP}}$ ($\mathrm{k_B}$/at) & $C_{11}$ (GPa) & $C_{12}$ (GPa)  & $C_{44}$ (GPa) & N$_{\mathrm{confs}}$\\
\hline
        AIMD  & -0.192 & -120.956 & 5.544 & 141 & 70 & 80 & 3546 \\
        EHCS             & -0.184 & -120.951 & 5.449 & 145 & 71 & 86 & 160  \\
        MLACS & -0.192 & -120.956 & 5.544 & 139 & 74 & 85 & 60   \\
\hline \hline
\end{tabular}
\end{table*}

\subsubsection{Bismuth at $\mathrm{T}=500~\mathrm{K}$}
The second {\it ab initio} example is the Bismuth in its rhomboedric phase at $\mathrm{T}=500~\mathrm{K}$ and $\mathrm{V}=34.547~\mathrm{\AA/at}$. This system is more anharmonic than the previous one, the temperature being close to the melting one ($\sim$ 550 K). A ($3\times3\times3$) supercell is built, with 128 atoms. The DFT calculations are performed using a norm-conserving pseudopotential, a cutoff energy equal to 15 Ha, a $(2\times2\times2)$ \textbf{k}-point mesh and with the generalized gradient approximation (GGA) Perdew–Burke–Ernzerhof (PBE) exchange-correlation functional. The time step used for the molecular dynamic simulations is $\Delta t \approx 3.6~\mathrm{fs}$.

\begin{figure}
    \centering
    \includegraphics[width=\linewidth]{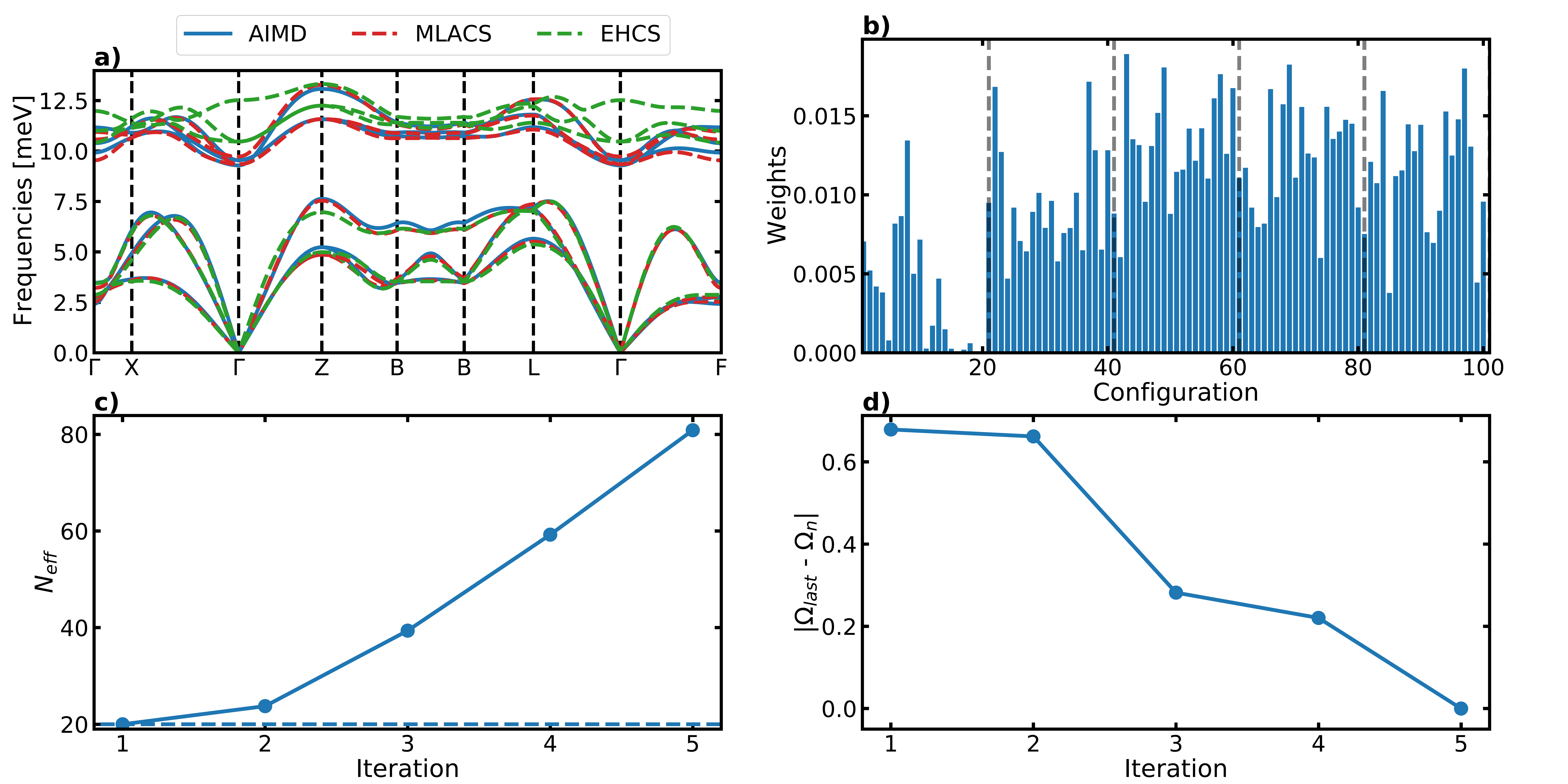}
    \caption{Same caption as Fig.~\ref{fig:Compare Si Tersoff} for Bismuth at $\mathrm{T}=500~\mathrm{K}$.}
    \label{fig:Compare Bi}
\end{figure}

\begin{figure}
    \centering
    \includegraphics[width=1.0\linewidth]{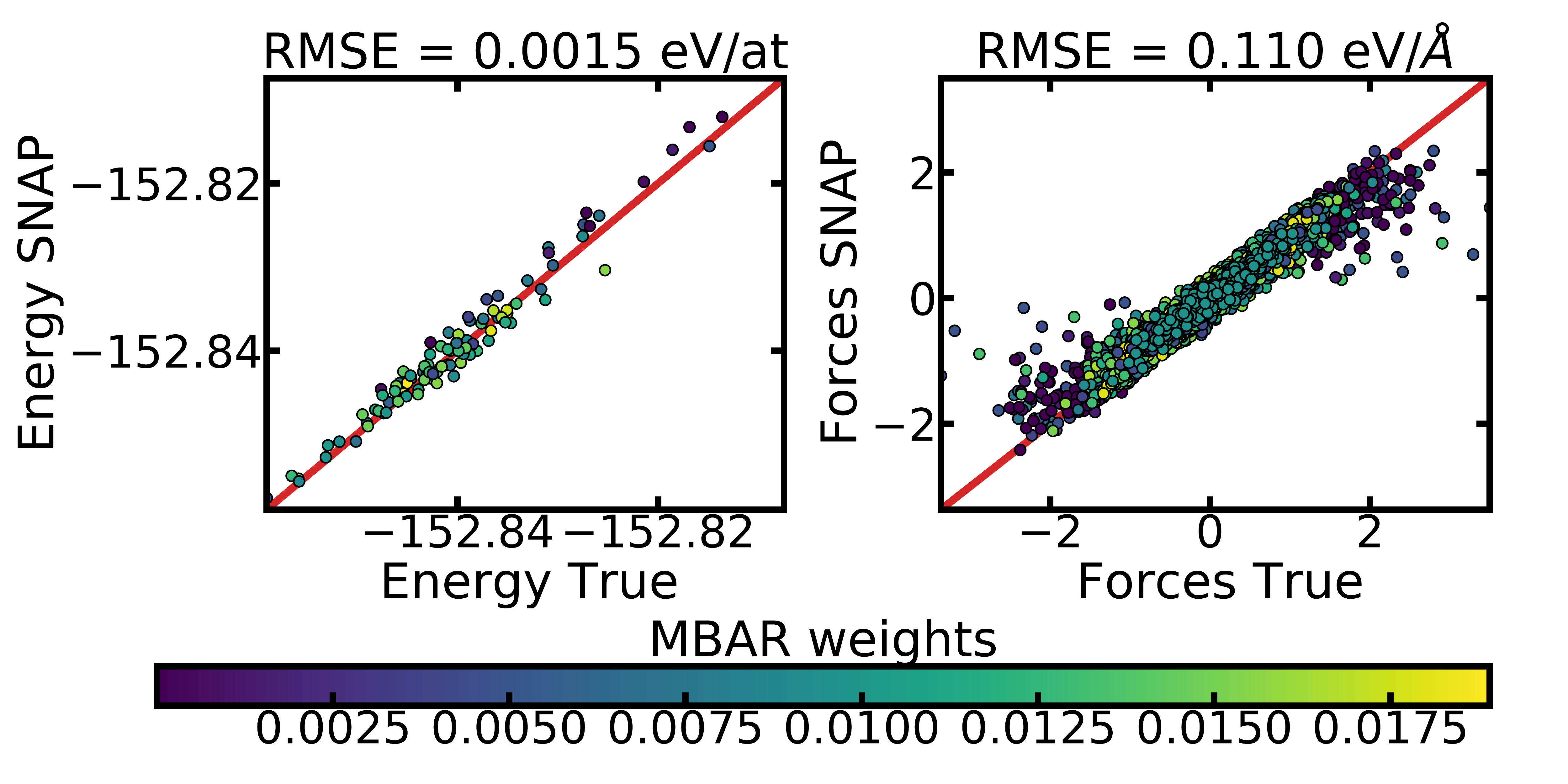}
    \caption{Same caption as Fig.~\ref{fig:Fit Silicium Tersoff} for Bismuth at $\mathrm{T}=500~\mathrm{K}$.}
    \label{fig:Fit Bismuth}
\end{figure}

All results are given in Fig.~\ref{fig:Compare Bi} and Tab.~\ref{tab:Compare Bi}. Those obtained by MLACS are in excellent agreement with AIMD (see also correlations in Fig.~\ref{fig:Fit Bismuth}) whereas EHCS leads to some discrepancies. As for Silicon, some optical branches are overestimated using EHCS (see the $\Gamma$ point), which can lead to significant errors when computing the thermal conductivity. Using MLACS, only 120 configurations are required, that is to say 30 times fewer configurations than in the AIMD trajectory (see Tab.~\ref{tab:Compare Bi}). The conclusions are the same as for Silicon: MLACS achieves a very good equilibrium canonical distribution in 10 hours, rather than 10 days using AIMD. 

\begin{table*}
\centering
    \caption{\label{tab:Compare Bi} Same caption as Tab.~\ref{tab:Compare Si Tersoff} for  Bismuth at $\mathrm{T}=500~\mathrm{K}$.}
\begin{tabular}{c c c c c c c c }
\hline \hline\\ [-0.3cm]
& $\FeTDEP^0$ (eV/at) & $\FeTDEP$ (eV/at) &  $\mathcal{S}_{\mathrm{TDEP}}$ ($\mathrm{k_B}$/at) & $C_{11}$ (GPa) & $C_{12}$ (GPa)  & $C_{44}$ (GPa) & N$_{\mathrm{confs}}$\\
\hline
        AIMD  & -0.234 & -153.143 & 8.445 & 73  & 21  & 30 & 3579 \\
        EHCS             & -0.231 & -153.134 & 8.363 & 69  & 16  & 32 & 240  \\
        MLACS & -0.234 & -153.140 & 8.447 & 70  & 25  & 32 & 120  \\
\hline \hline
\end{tabular}
\end{table*}

\subsubsection{MgO at $\mathrm{P}=300~\mathrm{GPa}$ and $\mathrm{T}=8000~\mathrm{K}$}
Here we study a system under extreme conditions: the B1 phase of MgO at $\mathrm{P}=300~\mathrm{GPa}$ ($\mathrm{V}=4.958~\mathrm{\AA/at}$) and $\mathrm{T}=8000~\mathrm{K}$. The supercell contains 64 atoms, the DFT calculations are performed using a projector augmented wave (PAW) atomic data, a cutoff energy equal to 30 Ha, a $(2\times2\times2)$ \textbf{k}-point mesh and a local density approximation (LDA) exchange-correlation functional. The time step used for the molecular dynamic simulations is $\Delta t \approx 1.2~\mathrm{fs}$.

\begin{figure}
    \centering
    \includegraphics[width=\linewidth]{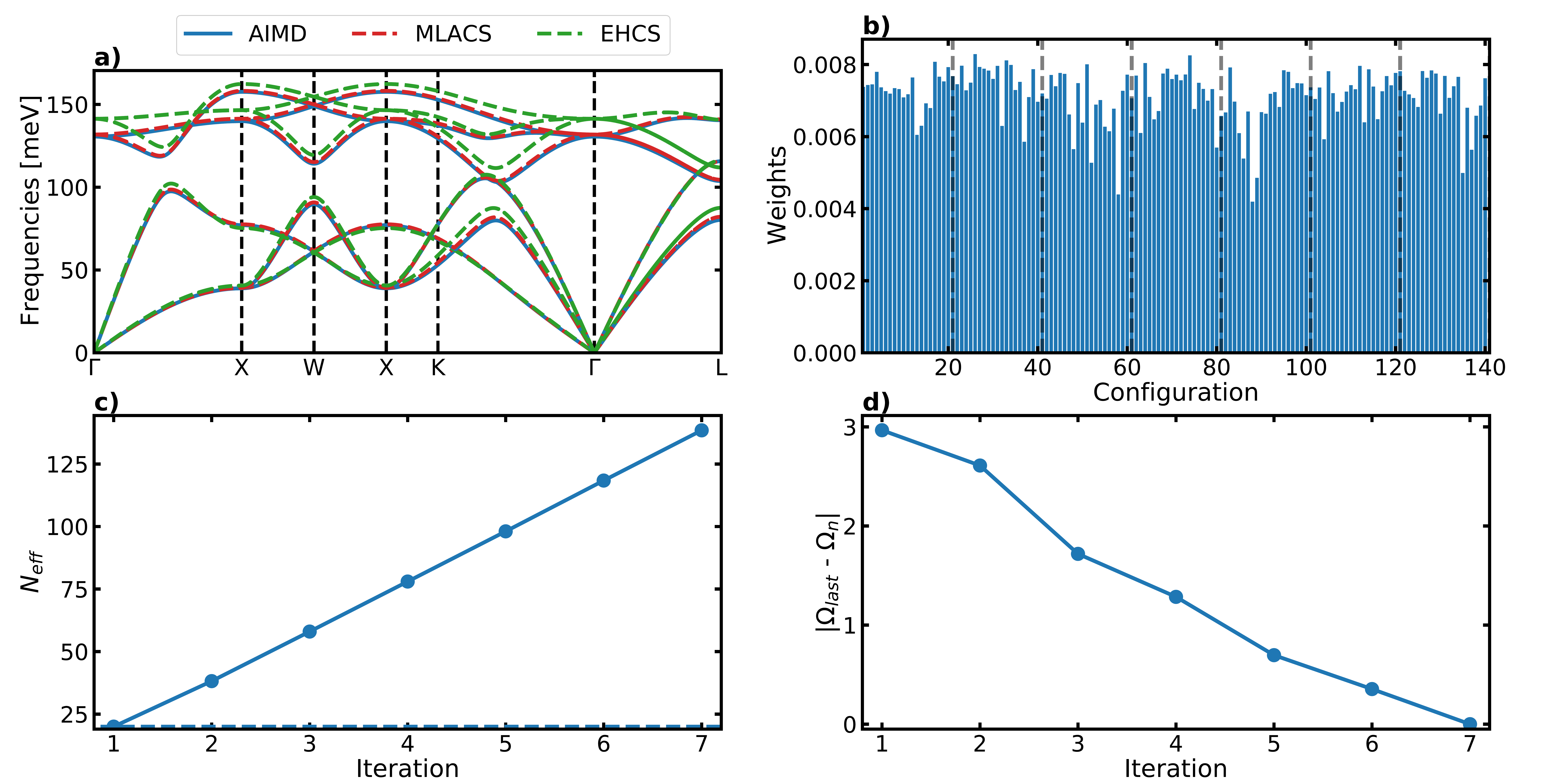}
    \caption{Same caption as Fig.~\ref{fig:Compare Si Tersoff} for MgO at $\mathrm{P}=300~\mathrm{GPa}$ and $\mathrm{T}=8000~\mathrm{K}$.}
    \label{fig:Compare MgO}
\end{figure}

\begin{figure}
    \centering
    \includegraphics[width=1.0\linewidth]{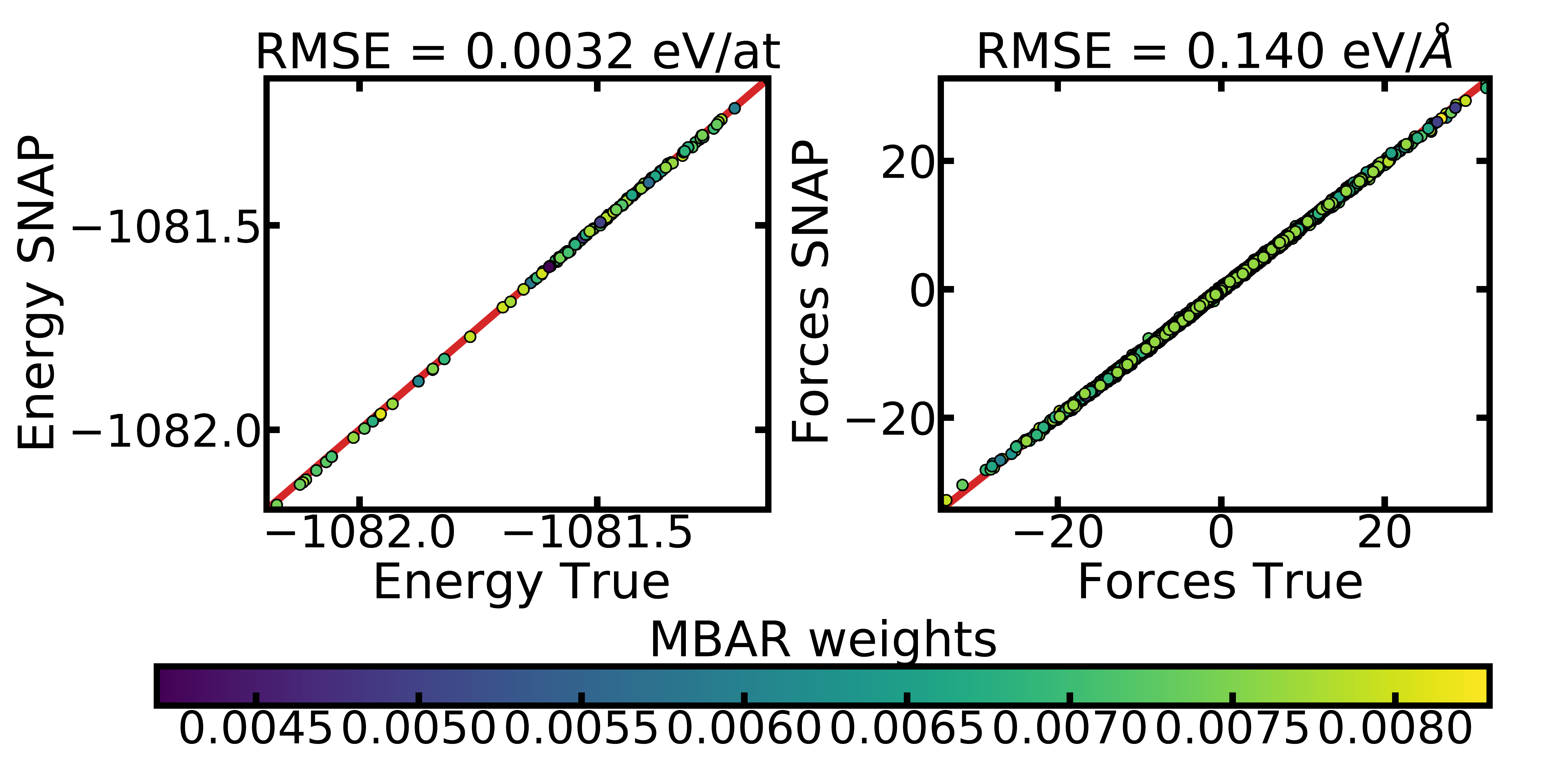}
    \caption{Same caption as Fig.~\ref{fig:Fit Silicium Tersoff} for MgO at $\mathrm{P}=300~\mathrm{GPa}$ and $\mathrm{T}=8000~\mathrm{K}$.}
    \label{fig:Fit MgO}
\end{figure}

Once again, the MLACS data are in very good agreement with the AIMD reference (see Fig.~\ref{fig:Compare MgO}, Fig.~\ref{fig:Fit MgO} and Tab.~\ref{tab:Compare MgO}). This example shows that, despite the high pressure (so high phonon frequencies) and high temperature, the MLACS method converges with an accuracy lower than 1 meV (see Fig.~\ref{fig:Compare MgO} d)). The free energy differences (using TDEP) are around 10 meV between MLACS and AIMD, compared to 100 meV between EHCS and AIMD. Moreover, the elastic properties are well reproduced using MLACS, with a difference around 1\%. A good description of these properties is crucial, this material being important in geophysics. The fast convergence is guaranteed by the reweighting performed at each iteration, which leads to a fast increase of the effective number of configurations (see Fig.~\ref{fig:Compare MgO} c)). In the end, only 160 DFT configurations are computed using MLACS, with respect to 7000 using AIMD. As previously, the computational time decreases from one week in AIMD to a few hours using MLACS.

\begin{table*}
    \caption{\label{tab:Compare MgO} Same caption as Tab.~\ref{tab:Compare Si Tersoff} for MgO at $\mathrm{P}=300~\mathrm{GPa}$ and $\mathrm{T}=8000~\mathrm{K}$.}
\begin{tabular}{c c c c c c c c }
\hline \hline\\ [-0.3cm]
& $\FeTDEP^0$ (eV/at) & $\FeTDEP$ (eV/at) & $\mathcal{S}_{\mathrm{TDEP}}$ ($\mathrm{k_B}$/at) & $C_{11}$ (GPa) & $C_{12}$ (GPa)  & $C_{44}$ (GPa) & N$_{\mathrm{confs}}$\\
\hline
        AIMD  & -4.182 & -1086.537 & 9.072 & 2651 & 675 & 174 & 7020 \\
        EHCS             & -4.091 & -1086.401 & 8.941 & 2828 & 528 & 189 & 720 \\
        MLACS & -4.163 & -1086.530 & 9.045 & 2682 & 675 & 176 & 160 \\
\hline \hline
\end{tabular}
\end{table*}

\subsubsection{Zirconium à $\mathrm{T}=1000~\mathrm{K}$}
We next consider Zirconium in its high temperature bcc $\beta$ phase, with a $(4\times4\times4)$ supercell containing 128 atoms, at $\mathrm{T}=1000~\mathrm{K}$ and $\mathrm{V}=22.945~\mathrm{\AA/at}$. This phase is unstable at low temperatures and the stabilisation at high temperatures comes from anharmonicity~\cite{Hellman_PRB84_2011,Bottin_CPC_2020,Anzellini_2020}. This peculiar behavior gives us the opportunity to test MLACS on a strongly anharmonic system. We use the PAW formalism, a cutoff energy equal to 15 Ha, a $(2\times2\times2)$ \textbf{k}-point mesh and the generalized gradient approximation (GGA) Perdew–Burke–Ernzerhof (PBE) exchange-correlation functional. The time step used for the molecular dynamic simulations is $\Delta t \approx 2.4~\mathrm{fs}$.

\begin{figure}
    \centering
    \includegraphics[width=\linewidth]{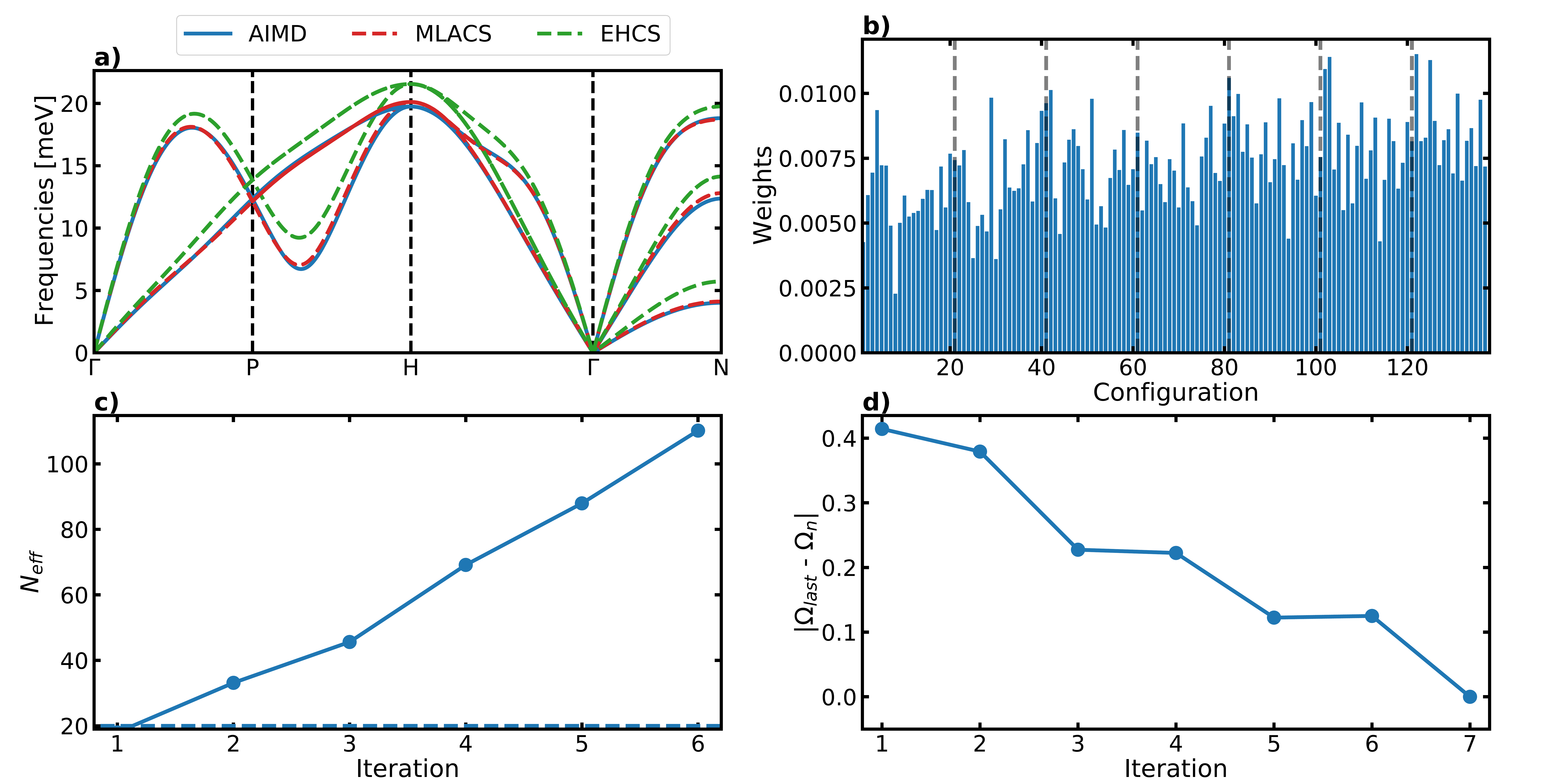}
    \caption{Same caption as Fig.~\ref{fig:Compare Si Tersoff} for Zirconium at $\mathrm{T}=1000~\mathrm{K}$.}
    \label{fig:Compare Zr}
\end{figure}

\begin{figure}
    \centering
    \includegraphics[width=1.0\linewidth]{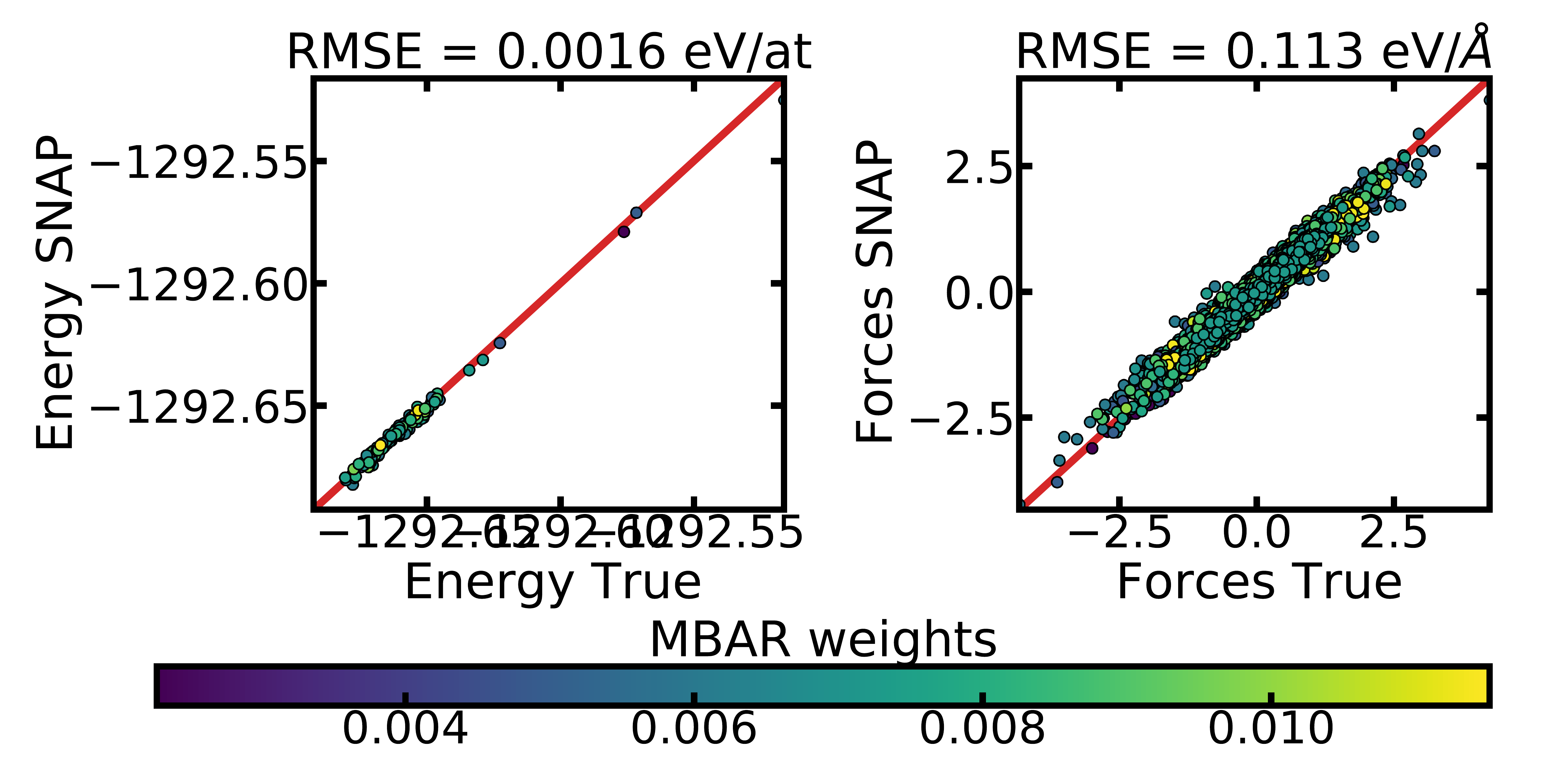}
    \caption{Same caption as Fig.~\ref{fig:Fit Silicium Tersoff} for Zirconium at $\mathrm{T}=1000~\mathrm{K}$.}
    \label{fig:Fit Zirconium}
\end{figure}

The results are shown in Fig.~\ref{fig:Compare Zr} and Tab.~\ref{tab:Compare Zr}. As previously we can see an overestimation of the phonon frequencies using EHCS with respect to AIMD. However, for this system, this occurs on acoustic branches, that is to say on the whole spectrum. Conversely, the phonon spectrum computed using MLCAS is in really good agreement with the AIMD one. Here, 160 configurations are generated using MLACS, with a reweighting leading to 100 effective configurations (see Fig.~\ref{fig:Compare Zr} c)), to compare to 7758 configurations in AIMD. Consequently, the acceleration is larger than previously. One DFT configuration taking about 15 minutes on 2000 processors, an AIMD simulation thus requires two months, where MLACS converges in two days. In Fig.~\ref{fig:Fit Zirconium} we show the correlations of forces and energies between the MLIP and DFT data.

\begin{table*}
    \caption{\label{tab:Compare Zr} Same caption as Tab.~\ref{tab:Compare Si Tersoff} for Zirconium at $\mathrm{T}=1000~\mathrm{K}$.}
    \begin{tabular}{c c c c c c c c }
\hline \hline \\ [-0.3cm]
& $\FeTDEP^0$ (eV/at) & $\FeTDEP$ (eV/at) & $\mathcal{S}_{\mathrm{TDEP}}$ ($\mathrm{k_B}$/at) & $C_{11}$ (GPa) & $C_{12}$ (GPa)  & $C_{44}$ (GPa) & N$_{\mathrm{confs}}$\\
\hline
        AIMD  & -0.528 & -1293.303 & 9.135 & 116 & 108 & 27 & 7758\\
        EHCS             & -0.497 & -1293.276 & 8.771 & 122 & 109 & 31 & 100 \\
        MLACS & -0.526 & -1293.298 & 9.109 & 118 & 110 & 28 & 160 \\
\hline \hline
\end{tabular}
\end{table*}

\subsubsection{Uranium at $\mathrm{T}=1200~\mathrm{K}$}
Finally, we study the bcc $\gamma$ phase of Uranium with a $(4\times4\times4)$ supercell containing 128 atoms, at $\mathrm{T}=1200~\mathrm{K}$ and $\mathrm{V}=20.618~\mathrm{\AA/at}$. As for $\beta$-Zr, $\gamma$-U is unstable at low temperatures and is stabilized at high temperatures by explicit temperature effects~\cite{Sderlind2012,Bouchet2017,Castellano2020}. We use the PAW formalism, a cutoff energy equal to 20 Ha, a $(2\times2\times2)$ \textbf{k}-point mesh and the GGA-PBE exchange-correlation functional. The time step used for the molecular dynamic simulations is $\Delta t \approx 3.6~\mathrm{fs}$.

\begin{figure}[b]
    \centering
    \includegraphics[width=\linewidth]{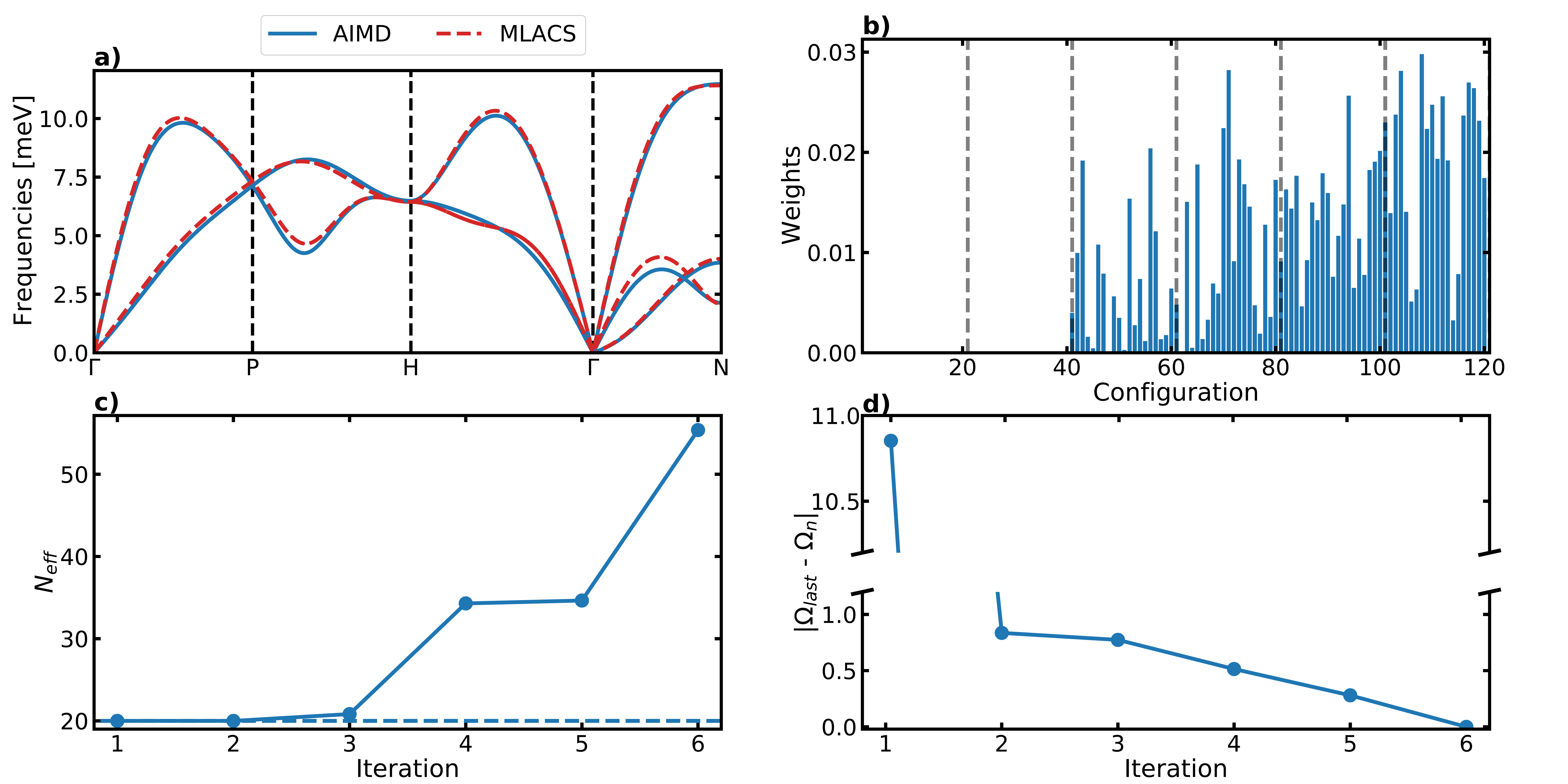}
    \caption{Same caption as Fig.~\ref{fig:Compare Si Tersoff} for Uranium at $\mathrm{T}=1200~\mathrm{K}$.}
    \label{fig:Compare U}
\end{figure}

\begin{figure}
    \centering
    \includegraphics[width=1.0\linewidth]{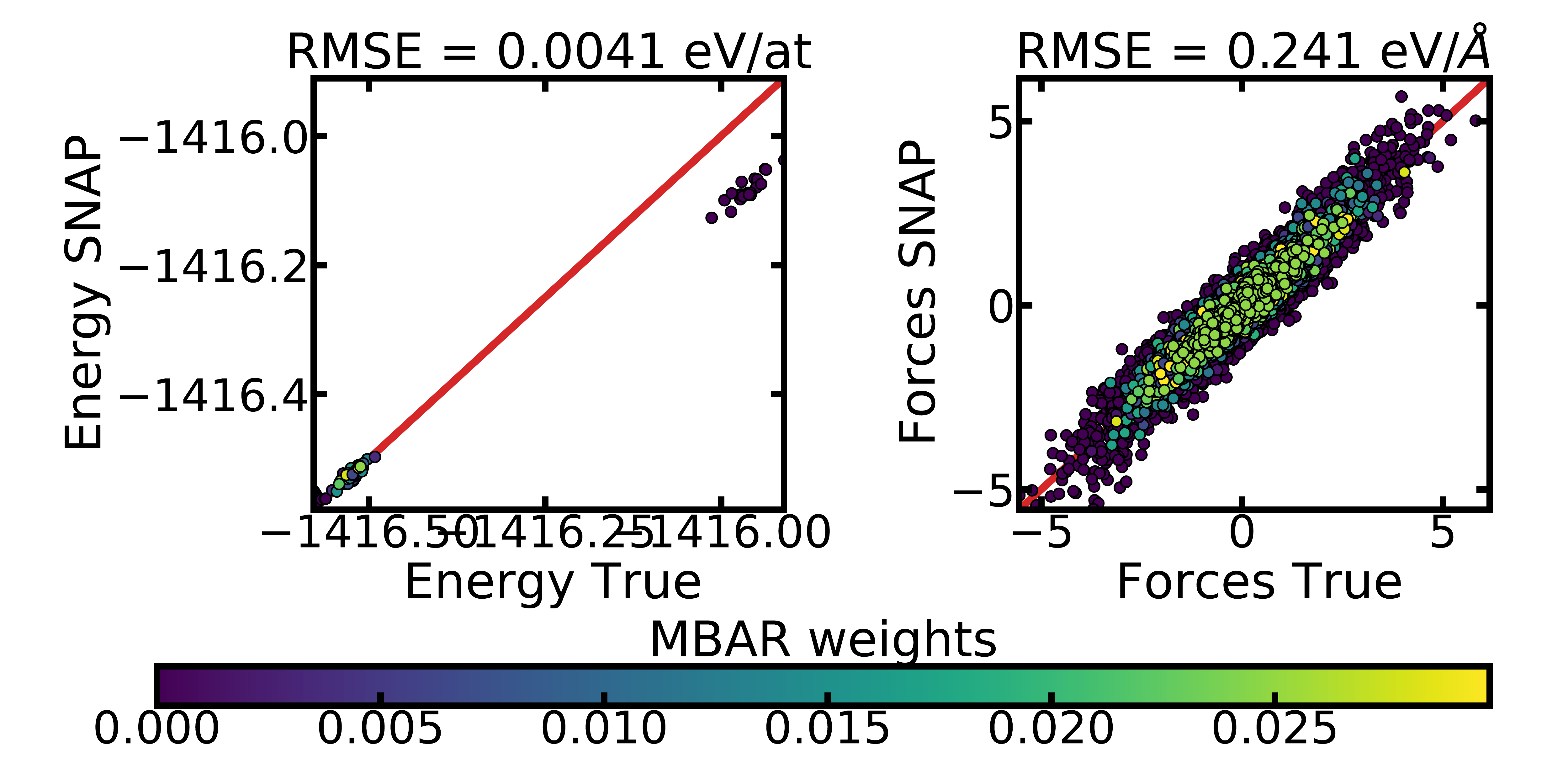}
    \caption{Same caption as Fig.~\ref{fig:Fit Silicium Tersoff} for Uranium at $\mathrm{T}=1200~\mathrm{K}$.}
    \label{fig:Fit Uranium}
\end{figure}

\begin{table*}
    \caption{\label{tab:Compare U} Same caption as Tab.~\ref{tab:Compare Si Tersoff} for Uranium at $\mathrm{T}=1200~\mathrm{K}$.}
    \begin{tabular}{c c c c c c c c }
\hline \hline \\ [-0.3cm]
& $\FeTDEP^0$ (eV/at) & $\FeTDEP$ (eV/at) & $\mathcal{S}_{\mathrm{TDEP}}$ ($\mathrm{k_B}$/at) & $C_{11}$ (GPa) & $C_{12}$ (GPa)  & $C_{44}$ (GPa) & N$_{\mathrm{confs}}$\\
\hline
        AIMD  & -0.894 & -1417.593 & 11.645 & 126 & 124 & 30 & 5981 \\
        MLACS & -0.887 & -1417.570 & 11.579 & 127 & 125 & 41 & 140  \\
\hline \hline
\end{tabular}
\end{table*}

This phase being highly anharmonic, the EHCS calculations never converged, despite several attempts and starting points (especially using the MLIP obtained after AIMD simulations). In Fig.~\ref{fig:Compare U} a) we show that the phonon spectrum obtained using MLACS reproduces correctly the AIMD one, even if some small discrepancies remain (lower than 1 meV). This slight disagreement leads to larger free energy differences than for the other examples we considered (see Tab.~\ref{tab:Compare U}). Once again, the reweighting using MBAR is crucial (see Fig.~\ref{fig:Compare U} b), c) et d)). The first iteration is far from the final result, with phonon frequency differences equal to 10 meV and an atomic structure which departs from the bcc phase and looks like a glass. However, from the second iteration on, the distribution gets closer to the final result and the phonon frequency differences strongly decrease. The 40 first configurations have a weight almost equal to zero. They are therefore discarded and not used in statistical averages (see also Fig.~\ref{fig:Fit Uranium}), whereas the next 80 configurations are included in the final equilibrium canonical distribution. This differs strongly from results for Si and $\beta$-Zr (see Fig.~\ref{fig:Fit Silicium} and Fig.~\ref{fig:Fit Zirconium}), for which all the configurations contribute. Despite this loss of data, MLACS still leads to a significant reduction of the computational cost: the AIMD trajectory performs 5981 time steps in two months, while MLACS converges using 140 configurations in two days.

\providecommand{\noopsort}[1]{}\providecommand{\singleletter}[1]{#1}%